\pdfoutput=1
\documentclass[fleqn,usenatbib]{mnras}

\usepackage{newtxtext,newtxmath}

\usepackage[T1]{fontenc}
\usepackage{ae,aecompl}

\usepackage{layouts}
\usepackage{graphicx}	
\usepackage{amsmath}	
\usepackage{amssymb}	

\newcommand{\mpc}{h^{-1}~\textrm{cMpc}}
\newcommand{\teff}{$\tau_\textrm{eff}$}

\newcommand{\rnf}{$R_{95}$}
\newcommand{\msol}{\textrm{M}_\odot}
\newcommand{\lya}{Ly-$\alpha$}
\newcommand{\deltaTeff}{$\delta_{\tau_\textrm{eff}}$}
\def\HI{\hbox{H$\,\rm \scriptstyle I$}}
\def\HII{\hbox{H$\,\rm \scriptstyle II$}} 

\def\HeII{\hbox{He$\,\rm \scriptstyle II$}}

\newcommand{\appropto}{\mathrel{\vcenter{
  \offinterlineskip\halign{\hfil$##$\cr
    \propto\cr\noalign{\kern2pt}\sim\cr\noalign{\kern-2pt}}}}}

\title[Characterising protoclusters with Lyman-$\alpha$ absorption]{Searching for the shadows of giants: characterising protoclusters with line of sight Lyman-$\alpha$ absorption} 

\author[J. S. A. Miller, J. S. Bolton and N. Hatch]{
Joel S. A. Miller\thanks{E-mail: joel.miller@nottingham.ac.uk},
James S. Bolton 
and
Nina Hatch
\\
School of Physics and Astronomy, University of Nottingham, University Park, Nottingham NG7 2RD, UK\\\\
}

\pubyear{2019}

\begin{document}
\label{firstpage}
\pagerange{\pageref{firstpage}--\pageref{lastpage}}
\maketitle

\begin{abstract}
\noindent We use state of the art hydrodyamical simulations from the Sherwood, EAGLE and Illustris projects to examine the signature of $M_{\rm z=0}\simeq 10^{14}M_{\odot}$ protoclusters observed in \lya\ absorption at $z\simeq 2.4$. We find there is a weak correlation between the mass overdensity, $\delta_{\rm m}$, and the \lya\ effective optical depth relative to the mean, \deltaTeff, averaged over $15~\mpc$ scales, although scatter in the $\delta_{\rm m}$--\deltaTeff\ plane means it is not possible to uniquely identify large scale overdensities with strong \lya\ absorption.  Although all protoclusters are associated with large scale mass overdensities, most sight lines through protoclusters in a $\sim 10^{6}$ $\rm cMpc^{3}$ volume probe the low column density \lya\ forest.  A small subset of sight lines that pass through protoclusters exhibit coherent, strong \lya\ absorption on $15h^{-1}\rm\,cMpc$ scales, although these correspond to a wide range in mass overdensity.  Assuming perfect removal of contamination by \lya\ absorbers with damping wings, more than half of the remaining sight lines with $\delta_{\tau_{\rm eff}}>3.5$ trace protoclusters.  It is furthermore possible to identify a model dependent $\delta_{\tau_{\rm eff}}$ threshold that selects only protoclusters.  However, such regions are rare: excluding absorption caused by damped systems, less than 0.1 per cent of sight lines that pass through a protocluster have $\delta_{\tau_{\rm eff}}>3.5$, meaning that any protocluster sample selected in this manner will also be highly incomplete.  On the other hand, coherent regions of \lya\ absorption also provide a promising route for identifying and studying filamentary environments at high redshift.

\end{abstract}




\section{Introduction}\label{sec:Intro}

Galaxy clusters at redshift $z=0$ are the most massive virialised objects in the Universe, residing in dark matter haloes with $M_\textrm{z=0}\geq 10^{14}~\msol$. At higher redshifts, the progenitors of these objects -- which are referred to as protoclusters -- are not yet virialised and are spread out over a scale of $\sim20\mpc$ at $z\sim2$ \citep{Chiang2013AncientProto-clusters,Muldrew2015WhatProtoclusters}.   Protoclusters are some of the densest structures on this scale at $z\sim 2-3$ when the cosmic star formation rate density is greatest, quasar activity peaked, and massive galaxies assembled the majority of their mass \citep{Madau2014CosmicHistory}. Protoclusters can therefore be used to study the effects of high density environments on galaxy formation and evolution at this important epoch,
and can also be used constrain structure formation and cosmological models via their growth rate \citep{Kravtsov2012FormationClusters}. 

The detection of protoclusters presents a challenge, however. Unlike mature clusters, they lack a hot X-ray emitting intra-cluster medium \citep{Overzier2016TheProtoclusters}. Therefore, they are most commonly found through the presence of large concentrations of galaxies. This a challenging observational task; protoclusters are large and diffuse objects, so they present only a small density enhancement over the field. Furthermore, protocluster galaxies are still forming, so they do not lie on a well defined red sequence \citep{Gladders2000AAlgorithm}.

Current observational techniques for finding protoclusters fall into two main categories. The first of these is to use typical galaxies as tracers. This has been performed using both photometric \citep{Daddi2009TwoRedshifts, Chiang2014DiscoveryCosmos} and spectroscopic \citep{Steidel2005SpectroscopicRedshift, Cucciati2014DiscoveryVUDS, Lemaux2014VIMOS3.3, Chiang2015SurveyingHETDEX, Lemaux2017TheZ4.57} surveys. The most massive protoclusters are rare (with number density $<10^{-6}\,\textrm{cMpc}^{-3}$) and spatially extended.  Consequently, in order to find protoclusters using this method, large cosmic volumes need to be surveyed. Until recently, galaxy redshift surveys that are deep enough to probe galaxies at $z\gtrsim2$ only cover areas of $\lesssim 1$ deg$^2$, such as the COSMOS survey \citep{Scoville2007TheOverview}, resulting in only a few tens of protoclusters discovered through this method over the last decade. This is about to change with ongoing and forthcoming surveys such as the Hyper Suprime-Cam Subaru Strategic Program (HSC-SSP)\footnote{https://hsc.mtk.nao.ac.jp/ssp/} and the Large Synoptic Survey Telescope (LSST)\footnote{https://www.lsst.org}; using only a small subset of the HSC-SSP data, \citet{Toshikawa2018GOLDRUSH.Area} have already discovered over 179 protocluster candidates.

The second of these methods is to focus searches around biased tracer objects, such as high redshift radio galaxies \citep{LeFevre1996Clustering3.14, Venemans2007I.Protoclusters, Hatch2011GalaxyGalaxies, Cooke2014AEnvironments}, quasars -- or quasi-stellar objects (QSOs) -- \citep{Wold2003OverdensitiesQuasars, Zheng2006AnQuasar, Capak2011AZ5.3, Wylezalek2013GalaxySpitzer, Morselli2014PrimordialLBT, Adams2015DiscoveryTelescope} and \lya\ blobs \citep{Badescu2017DiscoveryZ=2.3, Cai2017Discovery2.3}, which are thought to correlate well with high density regions at high redshift, and/or be the progenitors of local brightest cluster galaxies. These are also very bright objects, and can therefore be found using significantly shallower surveys. This approach greatly reduces the observational cost and it is how the majority of presently spectroscopically confirmed protoclusters at $z \gtrsim 1.3$ have been discovered. However, this comes at the expense of introducing a strong selection bias to the resulting sample.

Protoclusters are not only traced by overdensities of galaxies, however, but also by intergalactic hydrogen gas \citep{Adelberger2003GalaxiesOverview, Mukae2017CosmicField}. The residual neutral hydrogen (H\,\textsc{i}) in this otherwise highly ionised medium can be detected in absorption in background quasar spectra through the resulting \lya\ absorption (i.e. the \lya\ forest).  The idea of locating protoclusters as regions overdense in H\,\textsc{i} was first proposed by \citet{Francis1993SuperclusteringTwo} with the first observational detection by \citet{Francis1996A2.38}.  Recently there has been renewed interest in this approach following the advent of large QSO surveys, such as the Baryon Oscillation Spectroscopic Survey (BOSS) \citep{Dawson2013TheSDSS-III}, the Extended Baryon Oscillation Spectroscopic Survey (eBOSS) and the forthcoming Dark Energy Spectroscopic Instrument (DESI) \citep{Vargas-Magana2019UnravelingDESI} and WEAVE-QSO surveys \citep{Pieri2016WEAVE-QSO:Telescope}. 

The recent work has been approached from two different directions. The first is \lya\ forest tomography \citep{Lee2014Observational2,Stark2015ProtoclusterMaps,Lee2018First2.55}, where the underlying 3-D mass distribution is reconstructed using multiple \lya\ forest spectra in the same patch of sky.  Applying insight obtained from a  $256^{3} \, h^{-3}\rm\,cMpc^{3}$ collisionless dark matter simulation, \citet{Lee2016ShadowField} have used this method to successfully detect a galaxy overdensity at $z=2.44$ in the COSMOS field, as well as cosmic voids at $z\sim2.3$ \citep{Krolewski2018DetectionField}.  The second method -- which we focus on in this work -- is to search for the most massive overdensities at $z=2\textrm{--}4$ from a large survey volume using strong, coherent \lya\ absorption along the line of sight \citep{Cai2016MAppingMethodology}.  This approach is more effective at redshifts where the density of bright background sources is too low for tomography. Guided by mock spectra extracted from a 1 $h^{-3}\rm\,cGpc^{3}$ collisionless dark matter simulation combined with an approximate scheme for modelling \lya\ absorption \citep{Peirani2014LyMAS:Field}, \citet{Cai2016MAppingMethodology} identified protoclusters as being closely associated with what they call Coherently Strong intergalactic \lya\ absorption systems (CoSLAs). Groups of CoSLAs have been used to discover a massive overdensity at $z=2.32$ \citep{Cai2017MAppingZ=2.32}. 

These approaches have had compelling success in locating potential protoclusters at $z\simeq 2-3$. However, fully hydrodynamical simulations that directly connect the gas distribution at $z>2$ to present-day clusters are required to establish whether coherent \lya\ absorption accurately tracks the progenitors of the largest collapsed clusters at $z=0$, or if a more complex relationship between mass and \lya\ absorption may disrupt this picture.  Furthermore, for protocluster searches that employ individual sight lines to detect CoSLAs, high column density \lya\ absorbers that possess large damping wings can be a significant contaminant \citep[cf.][]{Cai2016MAppingMethodology}. 

In this work we shall address these points by investigating the properties of \lya\ absorption systems within protoclusters at $z > 2.4$ using fully hydrodynamical simulations performed by the Sherwood \citep{Bolton2017The5}, \textsc{eagle} \citep{Schaye2015TheEnvironments} and Illustris \citep{Vogelsberger2014IntroducingUniverse} projects.   With a typical volume of order $10^{6}\rm\,cMpc^{3}$, these simulations only contain of order $\sim 10$ clusters with masses $M_{\rm z=0}\geq 10^{14}M_{\odot}$.  On the other hand, unlike larger collisionless dark matter simulations, they include a full treatment of the gas physics using a variety of different (sub-grid) feedback models.  More importantly, however, these simulations also have sufficient resolution to correctly incorporate \lya\ absorption features over a wide range of \HI\ column densities, including absorption systems with damping wings.  Our approach therefore provides a complementary perspective on the relationship between \lya\ absorption on small scales and the distribution of mass in protoclusters at $z\simeq 2.4$.  Furthermore, as we explicitly track the formation of structure to $z=0$ in the simulations, we are able to assess the completeness and contamination of a sample of protocluster candidates selected using coherent line of sight \lya\ absorption. 

This paper is structured as follows.  We first introduce the hydrodynamical simulations we use in Section~\ref{sec:Sims}, and compare the \HI\ column density distribution in each model to observational constraints in Section~\ref{sec:CDDF}. In Section~\ref{sec:Sample} we describe the basic properties of protoclusters in the simulations.  We characterise mass overdensities on large scales by their \lya\ absorption in Section~\ref{sec:Results}, and examine their connection to protoclusters in Section~\ref{ssec:PCFrac}.  Finally, the effectiveness of using line of sight \lya\ absorption to detect candidate protoclusters at $z\simeq 2.4$ is discussed in Section~\ref{sec:CompCont}, before we summarise and conclude in Section \ref{sec:Conclusion}.  Throughout the paper, we refer to comoving distance units using the prefix \lq\lq c".

\section{Hydrodynamical simulations}\label{sec:Sims}

The hydrodynamical simulations used in this work are summarised in Table \ref{tab:Sims}.  Outputs at two different redshifts were used for each model: the output closest\footnote{This corresponds to $z=2.4$ for Sherwood, $z=2.478$ for \textsc{EAGLE} and $z=2.44$ for Illustris.} to $z=2.4$, along with the corresponding output at $z=0$.  Below, we briefly describe the relevant properties of each simulation in turn, although further, extensive descriptions of these simulations may be found elsewhere.

\begin{table*}
	\centering
	 \caption{Hydrodynamical simulations used in this work.  The columns list, from left to right: the simulation name, the box size in $h^{-1}\, \rm cMpc$, the total number of resolution elements, the dark matter and gas particle masses (or equivalently for Illustris, the typical hydrodynamical cell mass), and the number of objects that have a total mass at $z=0$  (as defined by a friends-of-friends halo finder) in the range $M_{\rm z=0}\geq10^{14}\rm\, M_{\odot}$ (clusters), $10^{13.75}\leq M_{\rm z=0}/M_{\odot}<10^{14}$ (large groups) and $10^{13.5}\leq M_{\rm z=0}/M_{\odot}<10^{13.75}$ (small groups).}
    \begin{tabular}{c|c|c|c|c|c|c|c}
    	\hline
    	Name & Box size & $N_\textrm{tot}$ & $M_\textrm{dm}$ & $M_\textrm{gas}$ & $M_{\rm z=0}\geq10^{14}~\msol$& $10^{13.75}\leq \frac{M_{\rm z=0}}{M_{\odot}}<10^{14}$ & $10^{13.5}\leq \frac{M_{\rm z=0}}{M_{\odot}}<10^{13.75}$ \\
        		   &[$h^{-1}\ $cMpc]&  & [$\msol$] & [$\msol$] & Clusters & Large groups & Small groups  \\
        \hline
        Sherwood & 80 & $2\times1024^3$ & $5.07\times10^7$ & $9.41\times10^6$  & 29 & 23 & 46 \\
        \textsc{eagle} & 67.77 & $2\times1504^3$ & $9.70\times10^6$ & $1.81\times10^6$ & 10 & 15 & 33 \\
        Illustris & 75 & $3\times1820^3$ & $6.26\times10^6$ & $1.26\times10^6$ & 14 & 18 & 29 \\
        \hline
    \end{tabular}
   
    \label{tab:Sims}
\end{table*}

\subsection{Sherwood}\label{ssec:Sherwood}

The Sherwood project \citep{Bolton2017The5} consists of a set of large, high resolution simulations of the \lya\ forest performed using a modified version of the smoothed particle hydrodynamics (SPH) code \textsc{P-Gadget-3}, last described in \citet{Springel2005TheGADGET-2}.  In this work we predominantly use the 80-1024-ps13 simulation from \citet{Bolton2017The5}, which we shall refer to as \lq\lq Sherwood". This simulation was performed in a $80^{3}h^{-3}\ \textrm{cMpc}^3$ volume using the star formation and galactic outflow model developed by \citet{Puchwein2013ShapingWinds}.  This assumes a \citet{Chabrier2003GalacticFunction} initial mass function (IMF) and a wind velocity, $v_{\rm w}$, that is proportional to the galaxy escape velocity, such that the wind mass-loading scales as $v_\textup{w}^{-2}$. 

The ultraviolet (UV) background follows the spatially uniform \citet{Haardt2012RadiativeBackground} model, which quickly reionises the IGM at $z=15$.  This model assumes the hydrogen is optically thin and in photo-ionisation equilibrium, and has an \HI\ photo-ionisation rate of $\Gamma_{\rm HI}=9.6\times10^{-13}\rm\,s^{-1}$ at $z=2.4$.  Additionally, a small boost to the \HeII\ photo-heating rate, $\epsilon_{\rm HeII}=1.7\epsilon_{\rm HeII}^{\rm HM12}$, has been applied at $2.2<z<3.4$ to better match observational constraints on the IGM temperature during and after \HeII\ reionisation \citep{Becker2011DetectionMedium}.  The Sherwood models adopt a Planck 2013 consistent cosmology \citep{PlanckCollaboration2014PlanckParameters} with $\Omega_m=0.308,\ \Omega_\Lambda=0.692,\ \Omega_b=0.0481,\ \sigma_8=0.826,\ n_s=0.963\ \textrm{and}\ h=0.678$.

In addition to the fiducial simulation described above, we also use the 80-1024 simulation described in \citet{Bolton2017The5}. This does not follow a physically motivated star formation model, but is identical to our fiducial run in all other respects.  Instead, gas particles with density $\Delta=\rho/\langle \rho \rangle > 1000$ and temperature $T < 10^5\,$K are converted directly into collisionless star particles. This \lq\lq quick \lya"\ approach increases computational speed while having a minimal effect on the low column density absorbers in the \lya\ forest \citep{Viel2004InferringSpectra}. As this removes all the cold, dense gas in the simulation, the incidence of \HI\ absorbers with column densities $N_{\rm HI} \gtrsim 10^{17}\rm\,cm^{-2}$ will be underpredicted.  We include this approach here, however, as it will more closely approximate results from earlier work using post-processed dark matter simulations that have insufficient resolution for modelling dense gas.  We refer to this model as Sherwood ``Q\lya".  

\subsection{EAGLE}\label{ssec:EAGLE}

The \textsc{EAGLE} (Evolution and Assembly of GaLaxies and their Environments) simulation \citep{Schaye2015TheEnvironments, Crain2015TheVariations, McAlpine2016TheCatalogues} was performed with a customised version of \textsc{P-Gadget-3}, where the standard SPH approach has been modified following the \textsc{anarchy} scheme described by \citet{Schaye2015TheEnvironments}.  The simulation we use in this work is the Ref-L0100N1504 model.  This has a smaller box size than Sherwood, corresponding  to $67.77^{3}\ h^{-3}\ \textrm{cMpc}^3$, but with a factor of $\simeq 5$ better mass resolution (see Table \ref{tab:Sims}).  Star formation is modelled using the approach of \citet{Schaye2008OnSimulations} assuming a Chabrier IMF, while stellar feedback follows the stochastic, thermal scheme described in \citet{DallaVecchia2012SimulatingFeedback}.   In addition, EAGLE follows gas accretion onto black holes; feedback from active galactic nuclei (AGN) is included using the methodology of \citet{Booth2009CosmologicalTests}.   

The spatially uniform, optically thin UV background used in EAGLE follows \citet{Haardt2001ModellingCUBA}; this quickly reionises the hydrogen at $z=9$.  An additional $2\rm\,eV$ per proton of energy from \HeII\ reionisation is also added at $z\simeq 3.5$, resulting in gas temperatures at mean density that are $\sim 4000\rm\,K$ larger compared to Sherwood by $z=2.4$. The \HI\ photo-ionisation rate at $z=2.478$ in the \citet{Haardt2001ModellingCUBA} model is $\Gamma_{\rm HI}=1.4\times 10^{-12}\rm\,s^{-1}$, a factor of $1.5$ larger than the more recent \citet{Haardt2012RadiativeBackground} UV background used in Sherwood.  \textsc{EAGLE} assumes a $\Lambda$CDM cosmology consistent with \citet{PlanckCollaboration2014PlanckParameters}, where $\Omega_m=0.307,\ \Omega_\Lambda=0.693,\ \Omega_b=0.04825,\ \sigma_8=0.8288,\ n_s=0.9611 \ \textrm{and}\ h=0.677$.

\subsection{Illustris} \label{ssec:Illustris}

Finally, we also use the Illustris-1 simulation (herafter referred to as ``Illustris") in our analysis \citep{Vogelsberger2014IntroducingUniverse, Nelson2015TheRelease}.  Unlike Sherwood and \textsc{EAGLE}, Illustris is performed with the moving-mesh hydrodynamics code \textsc{arepo} \citep{Springel2010EMesh}.  Illustris has a slightly smaller volume than Sherwood ($75^{3}\ h^{-3}\ \textrm{cMpc}^3$), but it has the highest mass resolution of the three simulations we consider. The star formation, stellar feedback and AGN feedback models are described in detail by \citet{Vogelsberger2013APhysics}. Star formation in Illustris also uses a Chabrier IMF and is based upon the \citet{Springel2003CosmologicalFormation} model. The stellar feedback uses a variable winds approach, where the wind velocity, $v_\textup{w}$, is scaled to the local dark matter velocity dispersion.  

The spatially uniform UV background follows \citet{Faucher-Giguere2009AReionization}, which quickly reionises the hydrogen in the simulation at $z=10.5$.  This UV background model has a \HI\ photo-ionisation rate $\Gamma_{\rm HI}=6.1\times 10^{-13}\rm\,s^{-1}$ at $z=2.44$ -- a factor of $0.6$ smaller than \citet{Haardt2012RadiativeBackground} --  and produces gas temperatures at mean density around $1500\rm\,K$ lower at $z=2.4$ compared to Sherwood.  

Furthermore, unlike Sherwood and EAGLE, Illustris uses an on-the-fly prescription for the self-shielding of hydrogen from Lyman continuum photons, following the approach of \citet{Rahmati2013OnSimulations}.  As we will discuss below, incorporating self-shielding is necessary for correctly capturing the incidence of absorption systems with column densities $N_{\rm HI}\geq 10^{17.2}\rm\,cm^{-2}$ (i.e. absorption systems that are optically thick to Lyman continuum photons).  The Illustris simulations assume a WMAP-9 consistent cosmology \citep{Hinshaw2013Nine-yearResults}, with $\Omega_m=0.2726,\ \Omega_\Lambda=0.7274,\ \Omega_b=0.0456,\ \sigma_8=0.809,\ n_s=0.963\ \textrm{and}\ h=0.704$.  

\subsection{Generation of mock \lya\ absorption sight-lines} \label{ssec:spectra}

Mock \lya\ absorption spectra were extracted along sight lines drawn from the Sherwood simulation using the SPH interpolation scheme described by \citet{Theuns1998P3M-SPHForest}, combined with the Voigt profile approximation from \citet{Tepper-Garcia2006VoigtFunction}.  Each sight line consists of $1024$ pixels, and is drawn in a direction parallel to the simulation boundaries, starting from a position selected at random on the projection axis. A total of $30,000$ sight lines ($10,000$ along each projection axis) were extracted from Sherwood, corresponding to an average transverse separation of $0.8h^{-1} \rm \,cMpc$.  The transmitted flux in each pixel is given by $F = e^{-\tau}$, where $\tau$ is the \lya\ optical depth.

Sight lines were extracted with the same average transverse separation from EAGLE and Illustris, although there are some small differences in the methodology due to the different hydrodynamics schemes employed by these models.  For EAGLE, the $M_{4}$ cubic spline kernel \citep{Monaghan1985AProblems} used in the standard version of \textsc{P-Gadget-3} was replaced with the $C_{2}$ \citet{Wendland1995PiecewiseDegree} kernel when performing the SPH interpolation.  In Illustris there are no smoothing lengths, $h_{\rm i}$, associated with the hydrodynamic cells.  Instead, we assign these based on the volume, $V_{\rm i}$, of each Voronoi cell, where 
\begin{equation} h_{\rm i} = \left(\frac{3N_{\rm sph} V_{\rm i}}{4\pi}\right)^{1/3}, \end{equation}
and we adopt $N_{\rm sph}=64$ for the number of smoothing neighbours.  We furthermore set all star-forming hydrogen gas with $n_{\rm H}>0.13\rm\,cm^{-3}$ to be fully neutral in Illustris, correcting for the unphysical neutral hydrogen fractions produced by the sub-grid star formation model.   In addition, since Illustris already incorporates self-shielded hydrogen on-the-fly, the \citet{Rahmati2013OnSimulations} prescription for self-shielding was applied in post-processing to both Sherwood and EAGLE.

Lastly, in order to correct for the approximately factor of two uncertainty in the UV background \HI\ photo-ionisation rate $\Gamma_{\rm HI}$ \citep{Bolton2005The2-4}, we rescale the optical depths in each pixel of the mock spectra by a constant, such that the \lya\ forest effective optical depth obtained from all the mock sight lines, $\tau_\textrm{eff} = -\ln\left<F\right>$, where $\left<F\right>$ is the mean transmitted flux, matches observational constraints \citep{Theuns1998P3M-SPHForest, Lukic2014TheSimulations}.  We use the \teff\ measurements from \citet{Becker2013ASpectra} for this purpose.  At $z=2.4$, these data correspond to $\tau_{\rm eff}=0.20$.

\section{The \HI\ column density distribution function}\label{sec:CDDF}

\begin{figure}
	\includegraphics[width=\columnwidth]{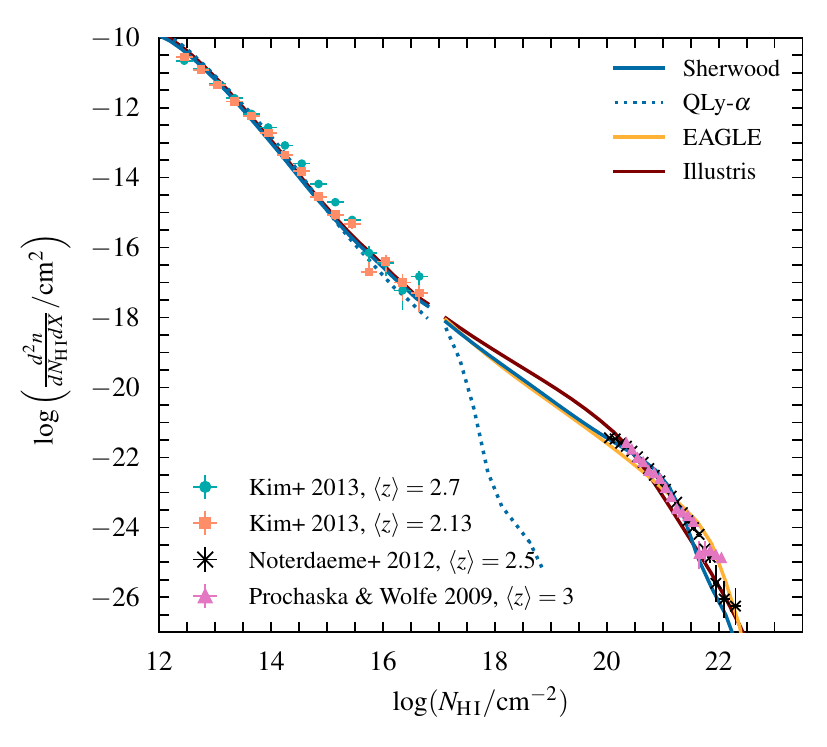}
	\vspace{-8mm}
    \caption{The \HI\ column density distribution function (CDDF) at $z\simeq 2.4$ obtained from Sherwood (solid blue curve), EAGLE (orange curve) and Illustris (brown curve). For comparison, the dotted blue line represents the CDDF from the Q\lya\ simulation.  Observational data from \citet{Kim2013The3.2} at $\left<z\right>=2.7$ and $\left<z\right>=2.13$ have been added at $N_{\rm HI}<10^{17}\rm\,cm^{-2}$, while data points from \citet{Noterdaeme2012Column9} at $\left<z\right>=2.5$ and \citet{Prochaska2009OnTime} $\left<z\right>=3$  are displayed at $N_{\rm HI}>10^{20}\rm\,cm^{-2}$.}
    \label{fig:CDDF}
\end{figure}

Before proceeding to analyse the properties of protoclusters in \lya\ absorption, we must first verify if the simulations reproduce the observed distribution of H\,\textsc{i} column densities\footnote{We will refer to \lya\ absorbers in four groups based on their column densities: \lya\ forest ($N_{\rm HI} < 10^{17.2}~\textrm{cm}^{-2}$), Lyman-limit systems (LLSs, $10^{17.2}\leq N_{\rm HI}/\textrm{cm}^{-2}<10^{19}$), super Lyman-limit systems (SLLSs, $10^{19}\leq N_{\rm HI}/\textrm{cm}^{-2} < 10^{20.3}$) and damped \lya\ absorbers (DLAs, $N_\textrm{HI}\geq 10^{20.3}~\textrm{cm}^{-2}$).  We will also refer to SLLSs and DLAs collectively as \lq \lq damped systems".}
at $z\simeq 2.4$.  The column density distribution function (CDDF) obtained from the three simulations are displayed Figure \ref{fig:CDDF}, along with observational measurements from \citet{Kim2013The3.2}, \citet{Noterdaeme2012Column9} and \citet{Prochaska2009OnTime}.  The simulated CDDFs at $N_{\rm HI}<10^{17}\rm\,cm^{-2}$ are calculated by integrating the \HI\ number density in each pixel in the mock sight lines over $50\rm\,km\,s^{-1}$ windows \citep{Gurvich2017TheForest}.  However, as absorption systems with $N_{\rm HI}>10^{17}\rm\,cm^{-2}$ are comparatively rare, we instead compute the CDDF by projecting the H\,\textsc{i} density for the entire simulation box onto a 2-D grid consisting of $20000^2$ pixels \citep{Altay2011ThroughSimulations,Rahmati2013OnSimulations,Bird2014DampedFeedback,Villaescusa-Navarro2018IngredientsMapping}.  The discontinuity seen at $N_{\rm HI}=10^{17}\rm\,cm^{-2}$ in Figure \ref{fig:CDDF} represents the point we switch from the integration method to the projection method. 

The fiducial Sherwood simulation (solid blue curve), as well as the EAGLE (orange curve) and Illustris (brown curve) are in good agreement with observational data over a wide range of column densities, although the level of agreement at $N_{\rm HI}>10^{21.3}\rm\,cm^{-2}$ is possibly fortuitous given that a treatment of molecular hydrogen is not included in our analysis \citep{Altay2013TheAbsorbers, Crain2017TheGalaxies}. The Illustris simulation also predicts a greater incidence of systems with $10^{17}\leq  N_\textrm{\rm HI}/\rm cm^{-2} \leq 10^{20}$, relative to EAGLE and Sherwood.  The reason for this difference is unclear, although we speculate this may be in part because the self-shielding correction is included on-the-fly in Illustris.   The photo-heating (and hence pressure smoothing) experienced by the high column density, self-shielded gas will therefore differ from the post-processed EAGLE and Sherwood runs.

For comparison, we also show the CDDF from the Q\lya\ model in  Figure \ref{fig:CDDF} (blue dotted curve). The Q\lya\ model is only a good match to the observational data for the \lya\ forest at $N_{\rm HI} \lesssim 10^{17}\rm\,cm^{-2}$. This is because most of the overdense gas that forms the strongest absorption systems has been converted into collisionless star particles (see Section \ref{ssec:Sherwood}).  Overall, this comparison demonstrates that Sherwood, EAGLE and Illustris will adequately capture the incidence of \lya\ absorbers at $z \simeq 2.4$ over a wide range of \HI\ column densities.

\section{Simulated protoclusters}\label{sec:Sample}
\subsection{Protocluster definitions}\label{ssec:defs}

\begin{figure*}
	\includegraphics[width=\textwidth]{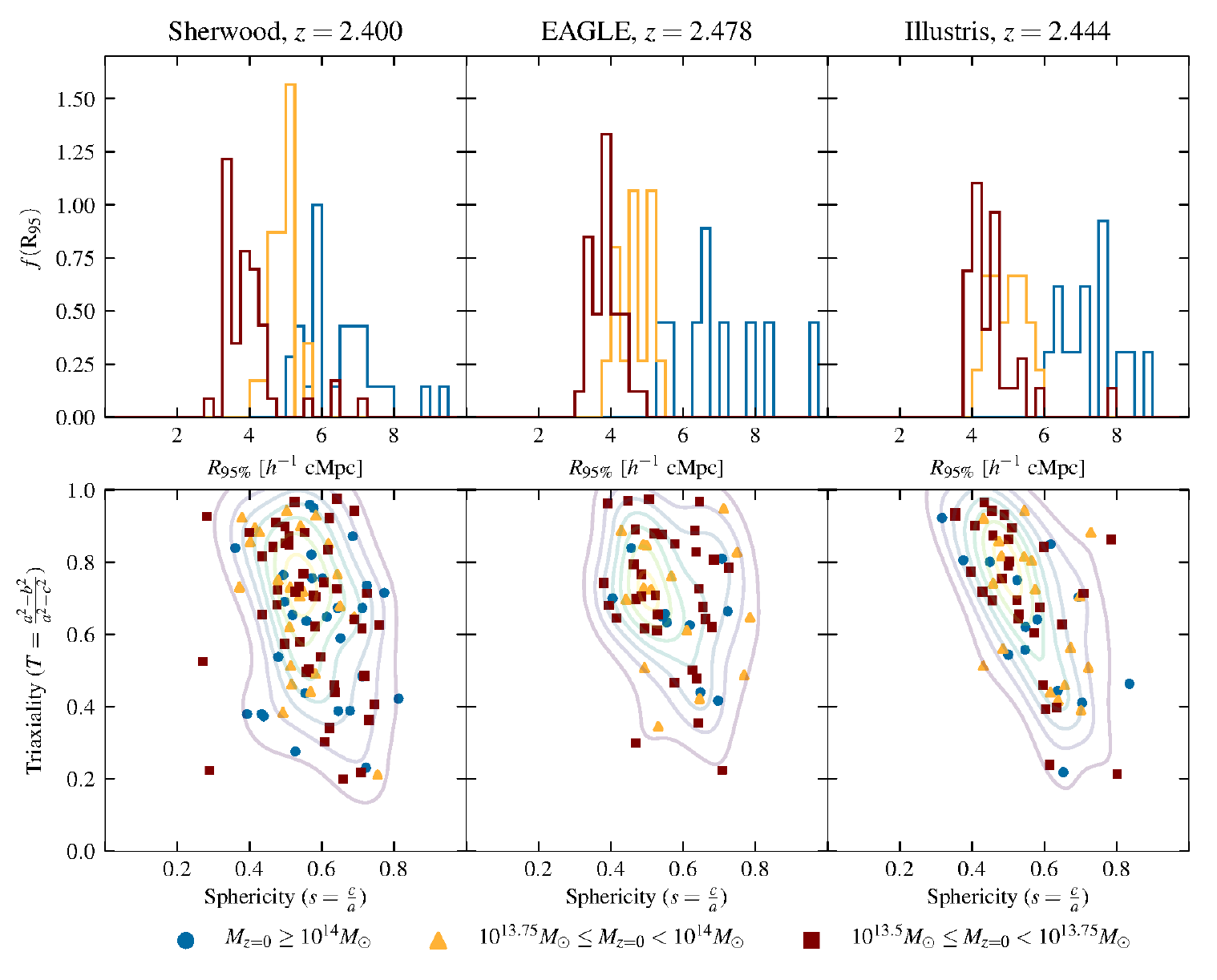}
	\vspace{-1.0cm}
    \caption{Top: Distribution of $R_{95}$ for the protoclusters (blue) and large and small protogroups (orange and brown) in Sherwood (left), EAGLE (centre) and Illustris (right) at $z\sim2.4$. Bottom: Scatter plots showing the triaxiality  (where $T=1$ and $T=0$ are prolate and oblate spheroids, respectively) against the sphericity (where $s=1$ and $s=0$ are spherical and aspherical, respectively) of the protoclusters and protogroups, where $a$, $b$ and $c$ are the principal semi-axes and $a\geq b\geq c$. Contours show the kernel density estimate for all points.}
    \label{fig:Summary}
\end{figure*}

\begin{figure*}
	\includegraphics[width=\textwidth]{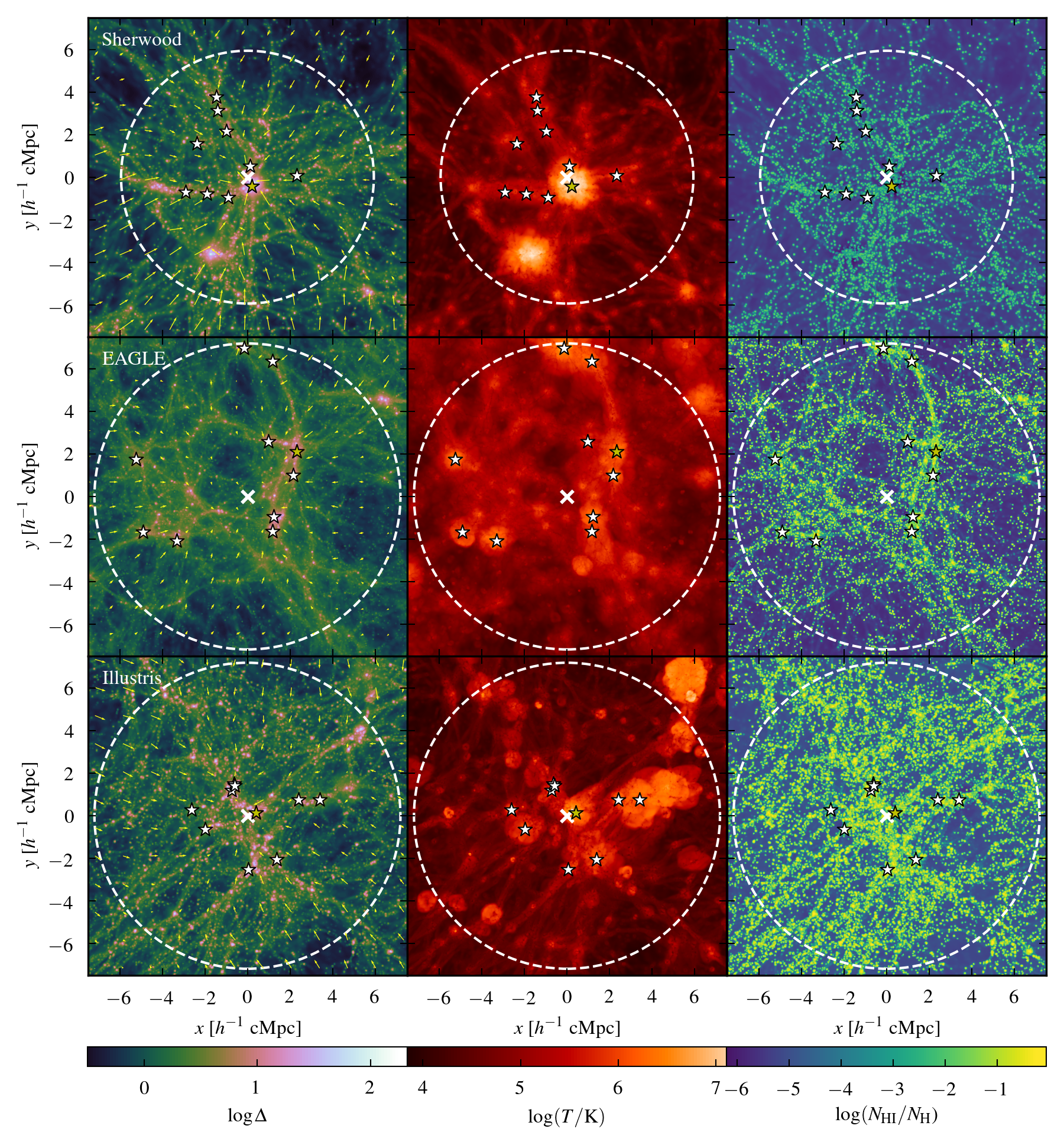}
	\vspace{-0.7cm}
    \caption{Projected maps of the normalised gas density, temperature and \HI\ fraction for protoclusters at $z\sim 2.4$ with masses $M_{\rm z=0}=10^{14.3}\,\msol$, $M_{\rm z=0}=10^{14.4}\,\msol$ and $M_{\rm z=0}=10^{14.3}\,\msol$ in Sherwood (top row), \textsc{eagle} (middle row) and Illustris (bottom row).  The projection depth along the $z$-axis is $15\mpc$, and is centred on the centre of mass of each protocluster. The dashed circles show $R_{95}$, and the star symbols correspond to the locations of the 10 most massive FoF groups within $R_{95}$.  The yellow stars correspond to the most massive FoF groups in each protocluster, which have mass $M=10^{13.6}\,\msol$, $M=10^{12.8}\,\msol$ and $M=10^{13.1}\,\msol$ in Sherwood, \textsc{eagle} and Illustris, respectively. The yellow arrows overlaid on the left column show the peculiar velocity field in the protoclusters relative to the centre of mass.}  
    \label{fig:3Panel}
\end{figure*}

\begin{figure*}
	\includegraphics[width=\textwidth]{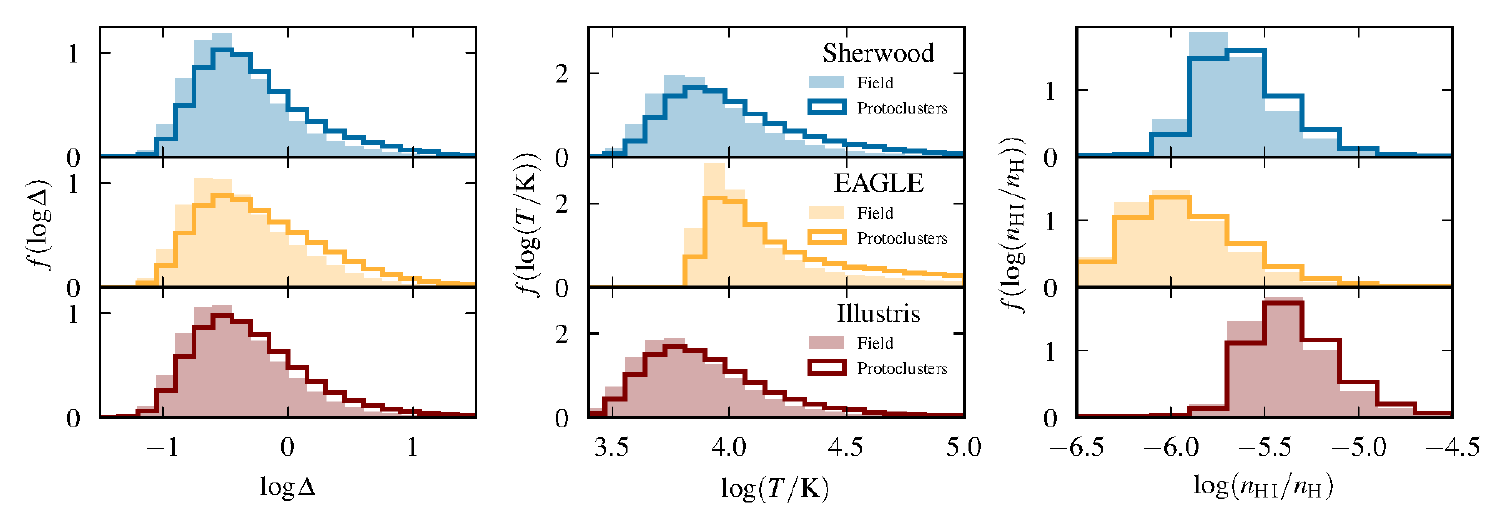}
	\vspace{-0.8cm}
    \caption{Volume weighted distributions of the normalised gas density, temperature and H\,\textsc{i} fraction in all protoclusters (line histograms) compared to the field (filled histograms) in  Sherwood (blue), EAGLE (orange) and Illustris (brown).}
    \label{fig:SimHist}
\end{figure*}

We now turn to describe the protoclusters in the simulations at $z\sim2.4$.  We identify the protoclusters by tracking all particles in a friends-of-friends (FoF) group\footnote{Note this means that our quoted protocluster masses will be systematically larger than virial mass estimators such as $M_{200}$ -- the total mass enclosed within a sphere whose mean density is $200$ times the critical density.  For example, \citet{White2000TheHalo} demonstrate that FoF halo masses will be approximately 10 per cent greater than $M_{200}$.} at $z=0$.  We will refer to the total mass of the $z=0$ FoF group as the mass of the protocluster, $M_{z=0}$.  Earlier studies \citep[e.g.][]{Muldrew2015WhatProtoclusters} have instead identified simulated protoclusters by tracing the merger tree of $z=0$ haloes back to the redshift of interest. However, as we are primarily interested in following large scale \lya\ absorption, in this work we also choose to follow the mass that is not bound in haloes at $z\simeq 2.4$.  

We consider three mass bins in our analysis: $M_\textrm{z=0}\geq10^{14}~\msol$ (clusters), $10^{13.75}~\msol\leq M_\textrm{z=0}<10^{14}~\msol$ (large groups) and $10^{13.5}~\msol \leq M_\textrm{z=0}<10^{13.75}~\msol$ (small groups).  We refer to the cluster progenitors as protoclusters, and the group progenitors as protogroups; we include the latter to provide a comparison to lower mass systems.  The number of FoF groups in each bin is summarised in Table \ref{tab:Sims}.  We also define the size of each of the protoclusters/groups at $z\simeq 2.4$ using \rnf,\, which corresponds to the radius of a sphere around the protocluster centre of mass that contains 95 per cent of $M_{\rm z=0}$.  

Additionally, following an analysis of the Millennium simulation \citep{Springel2005SimulationsQuasars} by \citet{Lovell2018CharacterisingProtoclusters}, for each protocluster we calculate two parameters derived from the principal semi-axes of a triaxial mass distribution, where $a\geq b\geq c$. The first is the axis ratio $s=c/a$ which provides a measure of sphericity, with $s=1$ corresponding to a spherical distribution and $s\sim0$ corresponding to a highly aspherical distribution. The second is the triaxiality parameter \citep{Franx1991TheShapes} given by 
\begin{equation}
	T=\frac{a^2-b^2}{a^2-c^2}
\end{equation}
This quantifies whether the mass distribution resembles a prolate ($T\sim 1$) or oblate ($T\sim0$) spheroid. 

\subsection{The size and shape of simulated protoclusters}

The \rnf\ distributions in the three mass bins are shown in the top row of Figure \ref{fig:Summary} for each simulation.  The protoclusters tend to be larger than the protogroups, and typically have $R_{95}=5$--$10\rm\,h^{-1}\,cMpc$.  These sizes are broadly consistent with the 90 per cent stellar mass radii recovered from the Millennium simulation at $z=2$ by \citet{Muldrew2015WhatProtoclusters}, as well as the radial extent determined by \citet{Hatch2011GalaxyGalaxies} for overdensities around radio galaxies at $z\sim2.4$.  The properties displayed in Figure \ref{fig:Summary} are also largely consistent between the different simulations, suggesting that (as expected on large scales) variations in the hydrodynamics schemes and sub-grid physics have little impact on the overall size and shape of the protoclusters.  Furthermore, in all three cases, the smaller, lower mass protogroups do not differ significantly in shape from the protoclusters.

As noted by \citet{Lovell2018CharacterisingProtoclusters}, however, a simple triaxial model does not fully capture the distribution of groups and filaments within protoclusters.  In Figure~\ref{fig:3Panel}, we therefore display two-dimensional projections of protoclusters selected from each of the three simulations: these have masses $M_{\rm z=0}=10^{14.3}\,M_{\odot}$ (Sherwood), $M_{\rm z=0}=10^{14.4}\,M_{\odot}$ (EAGLE) and  $M_{\rm z=0}=10^{14.3}\,M_{\odot}$ (Illustris).  Several different protocluster morphologies are apparent, ranging from a structure dominated by a massive central halo (Sherwood) to a more diffuse structure with multiple, lower mass haloes (e.g. EAGLE).  Figure~\ref{fig:3Panel} also shows the location of the most massive FoF group (yellow stars) along with the next nine most massive FoF groups (white stars) in each protocluster.  These are located in overdense regions where, in general, the gas temperatures are higher as a result of shock heating and feedback from stellar winds and/or black hole accretion. It is also apparent that regions with high H\,\textsc{i} fractions trace the overdense gas; photo-ionisation equilibrium with the UV background means the \HI\ number density, $n_{\rm HI}$, scales with the square of the gas density.

Finally, in Figure \ref{fig:SimHist} we show the volume weighted distribution of the gas density, temperature and \HI\ fraction for the protoclusters compared to the ``field" (i.e. regions outwith $R_{95}$) in each of the simulations. The differences in gas properties between the protoclusters and the field are small, but as might be anticipated from an inspection of Figure~\ref{fig:3Panel}, the protoclusters are slightly more dense, hotter and have a larger \HI\ fraction. Figure \ref{fig:SimHist} also highlights the differences in gas properties between the three different simulations. Although the gas density distribution is similar in all three cases, there are differences in the temperature and \HI\ fraction distribution that arise from the different UV background models (see Section~\ref{sec:Sims}).  Illustris (EAGLE) has a slightly larger (smaller) average H\,\textsc{i} fraction compared to Sherwood.  This is due to the smaller (larger) \HI\ photo-ionisation rate used in the UV background model and -- to a lesser extent -- the dependence of the \HI\ fraction on the lower (higher) gas temperature through the \HII\ recombination rate.

\section{Characterising mass overdensities in \lya\ absorption on $15\mpc$ scales}\label{sec:Results}

\begin{figure*}
	\includegraphics[width=\textwidth]{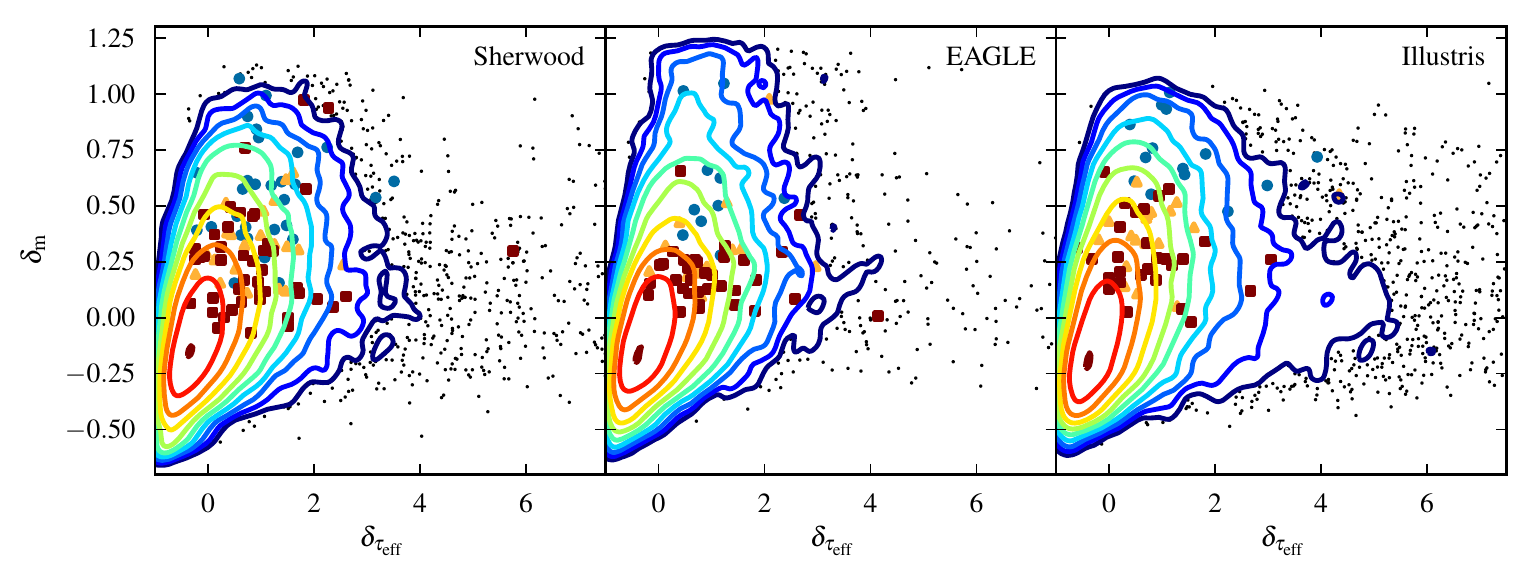}
	\vspace{-0.8cm}
    \caption{The relationship between $\delta_m$ and \deltaTeff\ on $15\,h^{-1}\rm\,cMpc$ scales \citep[following][]{Cai2016MAppingMethodology}. Left to right, the panels show the results from Sherwood, EAGLE and Illustris. The contours represent the number density of data points relative to the central contour, with the number density decreasing by $1/3$ dex with each contour.  Black points represent every system which lies outside of these contours (i.e. with relative number density $<10^{-3}$). Additional points corresponding to the centre of mass of the protoclusters (blue circles) large protogroups (orange triangles) and small protogroups (brown squares) are also displayed.}
    \label{fig:ContourPS13_All}
\end{figure*}

\begin{figure*}
    \includegraphics[width=\textwidth]{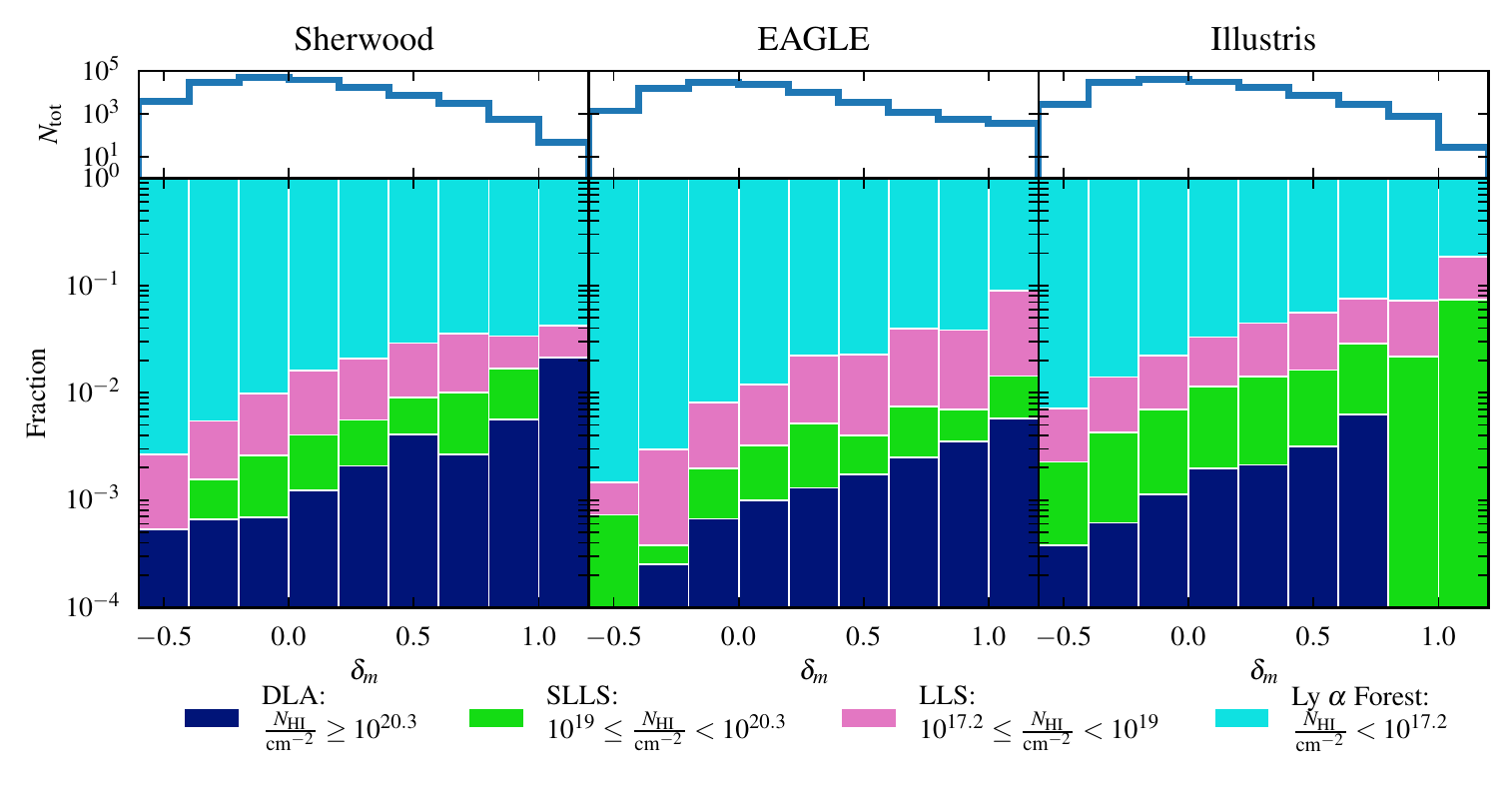}
    \vspace{-0.8cm}
    \caption{The fraction of $15\mpc$ segments in bins of $\Delta\delta_\textrm{m}=0.2$ whose largest constituent \HI\ column density measured on $50~\textrm{km s}^{-1}$ scales corresponds to either \lya\ forest, LLSs, SLLSs or DLAs in Sherwood (left), EAGLE (centre) and Illustris (right). The upper panels display the total number of $15\mpc$ segments in each bin.}
    \label{fig:MassNHIFrac}
\end{figure*}

Observationally, protoclusters are identified as large scale, high density regions. For this reason, we first investigate the \lya\ absorption properties of mass overdensities within the hydrodynamical simulations, with no consideration as to whether these eventually collapse to form a cluster by $z=0$.  From this, we determine the effectiveness of using  \lya\ absorption to detect large scale overdense regions at $z=2.4$.  It is important to note, however, that this does not address how effectively \lya\ absorption probes the gas at $z\simeq 2.4$ that \emph{actually} collapses to form clusters by $z=0$ -- we consider this further in Section~\ref{ssec:PCFrac}.

\subsection{Relationship between mass and effective optical depth on $15\,h^{-1}\rm\,cMpc$ scales}\label{ssec:MassVsTeff}

In order to assess how well \lya\ absorption traces large scale, high density regions, we consider the correlation between mass and \lya\ effective optical depth on $15\mpc$ scales. We choose this scale as it corresponds to the characteristic size of protoclusters at $z\geq2$ \citep{Chiang2013AncientProto-clusters, Muldrew2015WhatProtoclusters}, and also follows the scale adopted by \citet{Cai2016MAppingMethodology}.  

Our procedure throughout this work is as follows.  First, we sum the masses of all particles in  $15^{3}\,h^{-3}\rm\,cMpc^{3}$ cubic volumes along every simulated \lya\ forest sight line (i.e. 30,\,000 in total for Sherwood, each with length $80h^{-1}\rm\,cMpc$), and compute the mass overdensity, $\delta_{\rm m}=\frac{m-\left<m\right>}{\left<m\right>}$ in each volume. Hence, $\delta_{\rm m}$ is a 3D average, where $m$ is the total mass within each $15^{3}\,h^{-3}\,\rm cMpc^{3}$ volume, and $\langle m \rangle$ is the mass contained in a $15^{3}\,h^{-3}\,\rm cMpc^{3}$ volume with density equal to the mean density, $\langle \rho\rangle=\rho_{\rm crit}\Omega_{\rm m}(1+z)^{3}$, where
\begin{equation} 
    \left<m\right> =4.25\times 10^{14}\,M_{\odot} \left(\frac{h}{0.678}\right)^{-1} \left(\frac{\Omega_{\rm m}}{0.308}\right).  
\label{eq:mass}\end{equation}
 \noindent
Masses above (below) this threshold represent overdense (underdense) volumes on $15\,h^{-1}\rm\,cMpc$ scales.  Next, we associate every $\delta_{\rm m}$ with the \lya\ forest effective optical depth, $\tau_{\rm eff}=-\ln\langle F \rangle$, obtained from the $15\,h^{-1}\rm\,cMpc$ segments of simulated spectrum that pass directly through the centre of each $15^{3}\,h^{-3}\,\rm cMpc^{3}$ volume\footnote{That is, the average transmission for each 1D spectral segment, $\langle F \rangle=e^{-\tau_{\rm eff}}$, is obtained by averaging over all the pixels in a section of simulated spectrum with length $15\,h^{-1}\rm\,cMpc$.}.  We then compute $\delta_{\tau_\textrm{eff}}=\frac{\tau_\textrm{eff}-\left<\tau_\textrm{eff}\right>}{\left<\tau_\textrm{eff}\right>}$, where $\tau_\textrm{eff}$ is the effective optical depth measured in each $15h^{-1}\rm\,cMpc$ spectral segment, and $\left<\tau_\textrm{eff}\right>=0.20$ is obtained from \citet{Becker2013ASpectra} at $z=2.4$.  In this way, we associate $\delta_{\tau_\textrm{eff}}$ from every 1D $15h^{-1}\rm\,cMpc$ segment with $\delta_{\rm m}$ in the surrounding 3D $15^{3}\,h^{-3}\,\rm cMpc^{3}$ volume \citep[see also][]{Cai2016MAppingMethodology}.

The relationship between $\delta_m$ and \deltaTeff, averaged over $15\mpc$ scales, is displayed in Figure~\ref{fig:ContourPS13_All}. In Sherwood, EAGLE and Illustris the bulk of the simulation box corresponds to volumes close to the mean density, and there is a weak positive correlation between $\delta_m$ and \deltaTeff.   The centre of mass for all protoclusters and most protogroups (shown by the blue, orange and brown symbols) reside in overdense volumes, although note these are not usually associated with segments of high \lya\ opacity; the protocluster centre of mass will not always coincide with a halo or large value of \teff\ (cf.  Figure~\ref{fig:3Panel}). The differences between the three simulations are minimal at $\delta_{\tau_{\rm eff}}\lesssim 2$, suggesting that variations in numerical methodology have little impact on the properties of the low density gas on large scales.  By contrast, more significant differences between the simulations are apparent at $\delta_{\tau_{\rm eff}}>2$, corresponding to segments containing high column density, self-shielded absorption systems.  These systems reside in both overdense and underdense volumes.  The number of  $\delta_{\tau_{\rm eff}}>2$ segments is greatest in Illustris, which is the simulation with the largest number of absorption systems at $10^{17}\leq N_{\rm HI}/\rm cm^{-2}\leq 10^{20}$ (see Figure~\ref{fig:CDDF}).

Haloes and galaxies are biased tracers of mass overdensity, therefore we expect a correlation between the presence of the high column density systems -- responsible for the high \deltaTeff\ tail in Figure~\ref{fig:ContourPS13_All} -- and $\delta_m$. In Figure~\ref{fig:MassNHIFrac} the fraction of $15\mpc$ segments containing an absorption system, classed as either a DLA, SLLS, LLS or as \lya\ forest based on the largest constituent column density in each segment, is shown in bins of $\delta_m$ for each of the simulations.  In all three simulations, the vast majority of the volume -- irrespective of overdensity -- is traced by \lya\ forest absorption, with 97.1, 98.0 and 93.5 per cent of $15\mpc$ segments containing systems with a maximum column density of $N_{\rm HI}<10^{17.2}~\textrm{cm}^{-2}$ in Sherwood, EAGLE and Illustris, respectively.  At all $\delta_m$, Illustris has a greater fraction of segments containing SLLSs than the other simulations, with $\sim20$ per cent of all segments associated with volumes of $\delta_m>1$ containing a LLS or SLLS. Note, however, that DLAs in Illustris are only present in segments associated with volumes of $\delta_{\rm m}\leq 0.8$, implying that absorption systems in the most overdense volumes in Illustris have slightly lower \HI\ fractions relative to EAGLE and Sherwood.  This is broadly consistent with the \HI\ column density distribution, where Illustris contains an excess of absorption systems with $10^{17}\leq N_{\rm HI}/\rm cm^{-2}\leq 10^{20}$ relative to Sherwood and EAGLE. Figure~\ref{fig:MassNHIFrac} also demonstrates that damped systems with $N_{\rm HI}>10^{19}\rm\,cm^{-2}$ are present at all overdensities on $15\mpc$ scales, explaining the scatter with $\delta_{\rm m}$ in the high \deltaTeff\ tail observed in Figure~\ref{fig:ContourPS13_All}.  

This highlights the importance of self-consistently modelling the incidence of high column density \HI\ absorption systems when identifying overdense volumes using \lya\ absorption.  Damped systems produce large values of \deltaTeff\ regardless of large-scale environment, and so segments of high \deltaTeff\ will not uniquely probe overdense volumes. In addition to this, in all three simulations there is an increase in the fraction of segments containing systems with $N_{\rm HI} \geq 10^{19} \textrm{cm}^{-2}$ with increasing $\delta_m$. This implies that highly overdense volumes, such as those that may collapse to form a cluster by $z=0$, will contain a greater fraction of sight lines that pass through a damped system.

\subsection{CoSLAs: are they mass overdensities?}

\begin{figure}
    \includegraphics[width=\columnwidth]{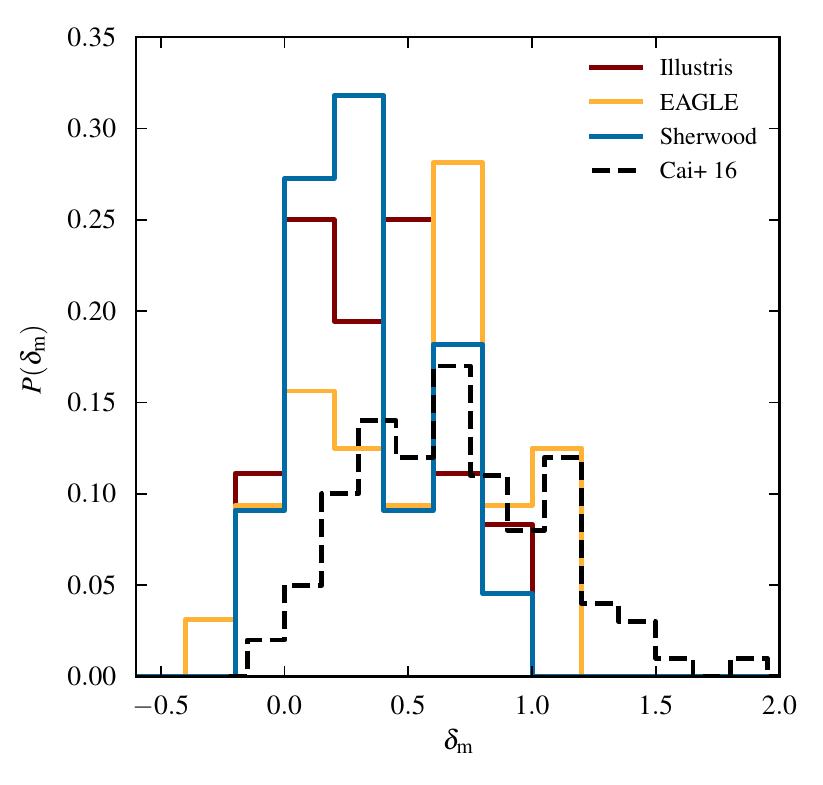}
    \vspace{-0.8cm}
    \caption{Probability distribution of $\delta_\textrm{m}$ for $15\mpc$ volumes with an associated $\delta_{\tau_\textrm{eff}} > 3.5$ (i.e. CoSLAs) after removing all sight lines containing damped systems in Sherwood (blue), \textsc{EAGLE} (orange) and Illustris (brown).  The corresponding distribution from the analysis of a $1\,h^{-3}\,\rm cGpc^{3}$ post-processed collisionless dark matter simulation by \citet{Cai2016MAppingMethodology} is displayed as a black dashed line. The total number of CoSLAs identified in each simulation is: Illustris (36), EAGLE (32), Sherwood (22).}
    \label{fig:CoSLAMassPDF}
\end{figure}

\begin{figure}
    \centering
    \includegraphics[width=1.1\columnwidth]{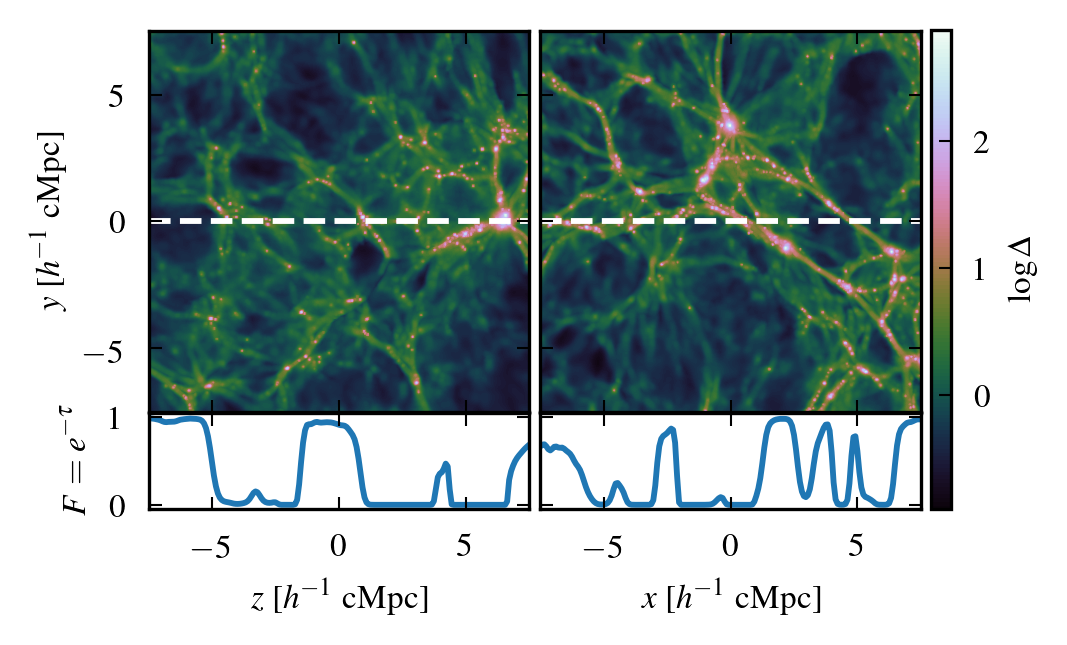}
    \vspace{-0.8cm}
    \caption{Examples of two $15\mpc$ volumes with an associated $\delta_{\tau_\textrm{eff}} > 3.5$ (i.e. CoSLAs) in Sherwood, one of which is drawn from an underdense volume with $\delta_\textrm{m}=-0.09$ (left), with the other corresponding to an overdense volume with $\delta_\textrm{m}=0.98$ (right).  In each case, the upper panels display the normalised gas density projected over a distance of $1\mpc$ centred on the sight line along which the \lya\ absorption spectrum is extracted (white dashed line), while the lower panels show the corresponding simulated \lya\ absorption. The underdense CoSLA has $\delta_{\tau_{\rm eff}}=3.76$ and a maximum $N_{\rm HI}=10^{18}~\textrm{cm}^{-2}$, while the overdense CoSLA has $\delta_{\tau_{\rm eff}}=3.68$ and a maximum $N_{\rm HI}=10^{15.3}~\textrm{cm}^{-2}$.}
    \label{fig:LowMassCont}
\end{figure}

From Figure~\ref{fig:MassNHIFrac}, it is evident that the majority of overdense volumes are traced by absorption from the \lya\ forest and LLSs.  We now explore whether any of these overdense volumes would be classified as being associated with Coherently Strong intergalactic \lya\ Absorption systems (CoSLAs) within the hydrodynamical simulations. \citet{Cai2016MAppingMethodology} (hereafter C16) associate protoclusters with CoSLAs, defining these as all segments of \lya\ absorption with $\delta_{\tau_\textrm{eff}}>3.5$ on $15\mpc$ scales after excluding any high column density absorbers with damping wings. Therefore, before selecting CoSLAs, we first remove all sight lines that contain column densities $N_{\rm HI}>10^{19}~\textrm{cm}^{-2}$ in the simulated spectra. We take this approach to ensure that we not only remove the segments contain the damped system itself, but also any neighbouring segments where extended damping wings from these systems are still present. Due to the correlation between $\delta_m$ and the fraction of damped systems (see Figure~\ref{fig:MassNHIFrac}), this process preferentially removes segments corresponding to overdense volumes.

\begin{figure}
    \includegraphics[width=\columnwidth]{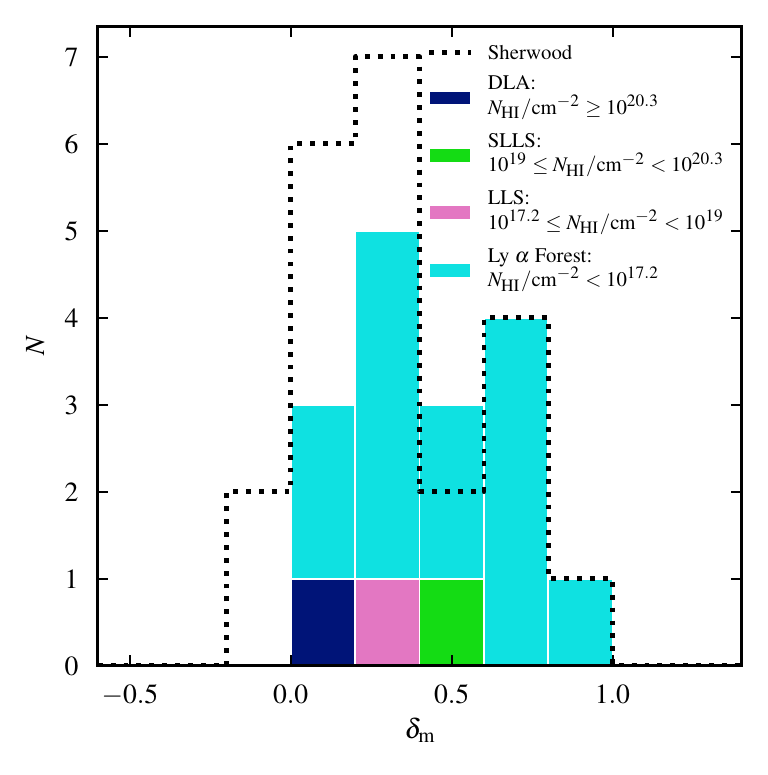}
    \vspace{-0.8cm}
    \caption{The filled histogram displays the number of $15\mpc$ volumes associated with segments of $\delta_{\tau_\textrm{eff}} > 3.5$ (i.e. CoSLAs) in bins of $\Delta\delta_\textrm{m}=0.2$ for the Q\lya\ simulation. The colour of each stacked bar indicates the class of absorber each segment corresponds to in Sherwood, which is identical to Q\lya\ except for the lack of any cold ($T<10^{5}\rm\,K$), dense ($\Delta>10^{3}$) gas.  The corresponding distribution for Sherwood is shown as the black dotted line, where there is a greater frequency of CoSLAs with $\delta_{\rm m}<0.4$. The total number of $15\mpc$ segments with $\delta_{\tau_\textrm{eff}} > 3.5$ in each simulation is: Sherwood (22), Q\lya\ (16).}
    \label{fig:QLyACoSLA}
\end{figure}

C16 used a collisionless dark matter simulation coupled with LyMAS \citep{Peirani2014LyMAS:Field} to show that almost all CoSLAs correspond to mass overdensities in the range $0<\delta_m<2$, with $\sim70$ per cent having $\delta_m\geq0.5$. \citet{Chiang2013AncientProto-clusters} showed that, at similar redshifts and scales ($z=2$ and $16.3h^{-1}~\textrm{cMpc}$ respectively), more than 80 per cent of volumes with $\delta_m\geq0.8$ will collapse to form clusters with $M_{z=0}>10^{14}~\msol$ by $z=0$.  The expectation is therefore that most CoSLAs will also probe structures that collapse to form clusters by $z=0$.  

In Figure~\ref{fig:CoSLAMassPDF}, we examine the $\delta_m$ associated with CoSLAs within each of the three hydrodynamical simulations. The distributions for Sherwood and Illustris are consistent with one another, with a median $\delta_m=0.36\pm0.07$ and $\delta_{\rm m}=0.35\pm0.08$, respectively, where we estimate the uncertainty by bootstrapping with replacement.  EAGLE has a higher median $\delta_{\rm m}=0.59\pm0.11$, formally differing from Illustris and Sherwood by $1.8\sigma$.  However, given the relatively small number of sight lines with coherent absorption on $15\mpc$ scales in the simulations, this may reflect differences in the rare, massive structures within each simulation.  Regardless of the median of each distribution, however, in all three simulations the CoSLAs cover a broad $\delta_m$ range including $\delta_m\leq0$, indicating that they do not exclusively probe large scale mass overdensities at $z=2.4$.  As a further illustration, Figure~\ref{fig:LowMassCont} displays examples of CoSLAs from both underdense ($\delta_m=-0.09$) and overdense ($\delta_m=0.98$) volumes in Sherwood.  In both cases, we observe that a CoSLA arises from the alignment of structure along the line of sight.  This can be due to LLSs associated with the alignment of several haloes punctuated by voids (left panel), or multiple lower column density absorbers associated with a more extended, filamentary structure (right panel).

These results differ from C16, who found CoSLAs to have a median $\delta_m=0.75\pm0.03$ and almost exclusively correspond to overdense volumes.  To explore the reason for this difference, we also consider the incidence of CoSLAs in the Q\lya\ simulation. The Q\lya\ model does not produce any damped systems and under produces LLSs (see Figure~\ref{fig:CDDF}) due to missing high density gas, but is otherwise identical to Sherwood.  Consequently, the Q\lya\ model should better approximate the lower resolution simulation\footnote{C16 use a collisionless dark matter simulation in a $1h^{-3}\rm\,cGpc^{3}$ volume with $1024^{3}$ particles, yielding a particle mass of around $9.6\times 10^{10}\rm\,M_{\odot}$.  For comparison, the typical dark matter particle mass needed to resolve the small scale structure of the \lya\ forest at $z\simeq 2$ is $\sim10^{7}M_{\odot}$ \citep{Bolton2009ResolvingSimulations}.} used in C16. 

In Figure~\ref{fig:QLyACoSLA}, the $\delta_m$ distribution of CoSLAs in Sherwood is compared to the Q\lya\ simulation (this time in terms of number in each bin).  The CoSLAs identified in Q\lya\ are coloured according to the maximum column density drawn from the matching locations within the Sherwood simulation.  There are several points to note from Figure~\ref{fig:QLyACoSLA}.  First, Q\lya\ contains two CoSLAs that are in segments containing SLLSs or DLAs in Sherwood (meaning these segments were discarded), while Sherwood contains eight segments classified as CoSLAs that are not present in Q\lya.  These eight CoSLAs all contain LLSs or high column density \lya\ forest systems in Sherwood, but have lower column densities -- and therefore lower values of \deltaTeff\ -- in Q\lya.  Furthermore, all these additional systems have $\delta_m<0.4$, which acts to dilute the correlation between $\delta_m$ and \deltaTeff.  This emphasises the importance of correctly capturing LLSs in the models; not only will a fraction of the CosLAs in the Q\lya\ model be contaminated by the presence of damped systems, but additional systems with $10^{16}\lesssim N_{\rm HI}/\rm cm^{-2}\leq 10^{19}$ are missed, resulting in an erroneously strong correlation between $\delta_m$ and \deltaTeff\ for CoSLAs. Second, the CoSLAs in Q\lya\ have a median $\delta_m=0.40\pm0.02$, with $\sim50$ per cent of CoSLAs having $\delta_m\geq0.5$.  This median $\delta_{\rm m}$ is consistent with the Sherwood simulation, although note it is still smaller when compared with C16.

\subsection{Effect of box size and mass resolution on CoSLAs}\label{ssec:ResBox}

Another important factor to consider is the box size (and therefore maximum cluster mass) and mass resolution used in the simulations.  We examine how mass resolution and box size impact on the relationship between mass overdensities and \deltaTeff\ on $15\mpc$ scales using four Q\lya\ runs drawn from the Sherwood simulation suite.  These simulations are summarised in Table~\ref{tab:ConvSims}.  As previously, sight lines were extracted with the same average transverse separation of $0.8h^{-1}\rm\, cMpc$ from all runs. 

\begin{table*}
	\centering
	    \caption{Hydrodynamical simulations used to test the effect of box size and mass resolution on CoSLAs.  All four simulations were performed using the quick-\lya\ approach that excludes dense, star forming gas (see Section~\ref{sec:Sims}).  The columns list, from left to right: the simulation name, box size, number of resolution elements, dark matter and gas particle masses, and the number of CoSLAs per unit volume obtained using sight lines with an average transverse separation of $0.8h^{-1}\rm\,cMpc$. All of the $80h^{-1}\rm\,cMpc$ boxes were performed with initial conditions generated using the same random seed.}
    \begin{tabular}{lcccccc}
      \hline
      Name & Box size         & $N_\textrm{tot}$ & $M_\textrm{dm}$ & $M_\textrm{gas}$  & $N_{\rm CoSLAs}/V$ \\
                 & [$h^{-1}\ $cMpc] &                  & [$\msol$]       & [$\msol$] &   [$ h^{3} \,\rm cMpc^{-3}$]\\
      \hline
      80-512 & 80 & $2\times 512^3$ & $4.06\times10^8$ & $7.52\times10^7$ &  $4.9\times10^{-5}$ \\
      80-1024 & 80 & $2\times 1024^3$ & $5.07\times10^7$ & $9.41\times10^6$ & $3.1\times 10^{-5}$\\
      80-2048 & 80 & $2\times 2048^3$ & $6.34\times10^6$ & $1.17\times10^6$ &  $1.4\times 10^{-5}$\\
      160-2048 & 160 & $2\times 2048^3$ & $5.07\times10^7$ & $9.41\times10^6$ &  $5.5\times 10^{-5}$\\
      \hline
    \end{tabular}
    \label{tab:ConvSims}
\end{table*}

\begin{figure}
    \centering
    \includegraphics[width=\columnwidth]{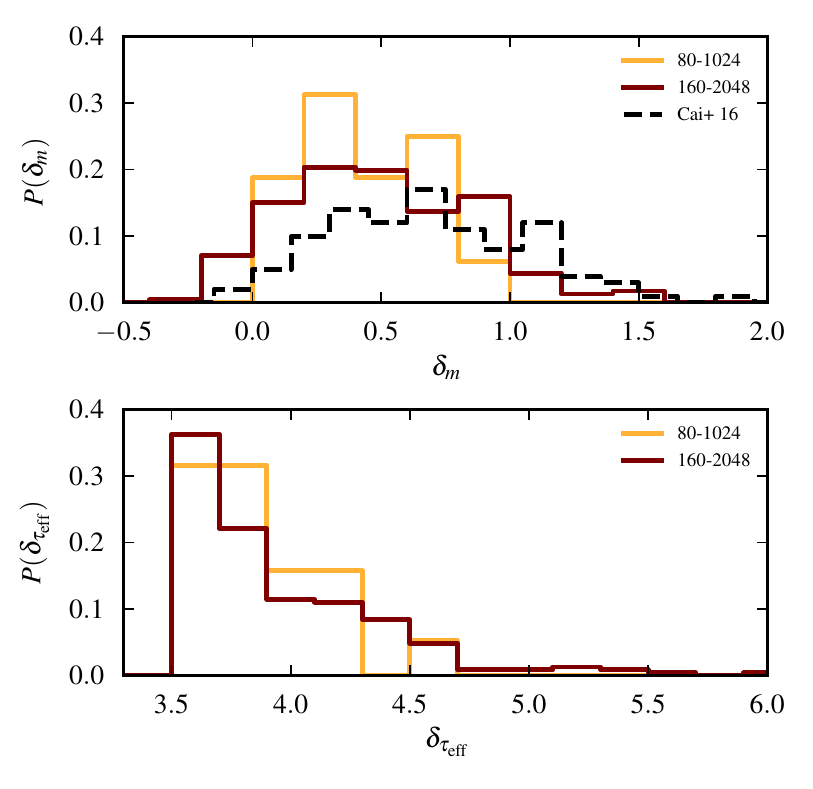}
    \vspace{-8mm}
    \caption{The probability distribution of CoSLAs as a function of $\delta_m$ (upper panel) and \deltaTeff\ (lower panel) for the Q\lya\ model (80-1024) compared to a simulation with the same mass resolution but a volume eight times larger (160-2048).  The $\delta_{\rm m}$ distribution from C16 is displayed as the dashed black histogram in the upper panel.} 
    \label{fig:CoSLA-BoxComp}
\end{figure}

\begin{figure}
    \centering
    \includegraphics[width=\columnwidth]{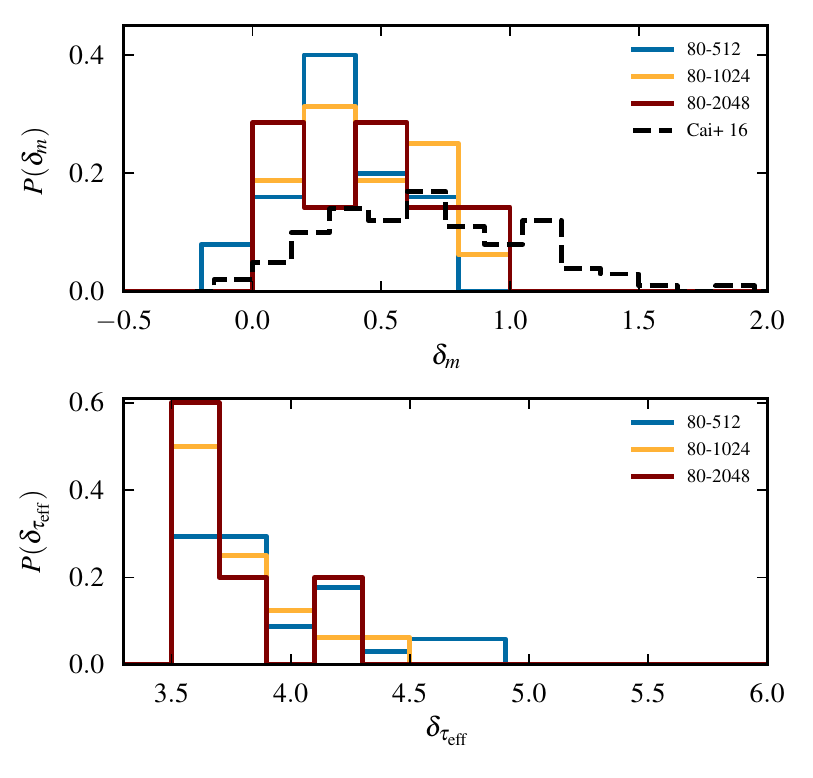}
    \vspace{-8mm}
    \caption{The probability distribution of CoSLAs as a function of $\delta_m$ (upper panel) and \deltaTeff\ (lower panel) for the Q\lya\ model (80-1024) compared to a simulations with the same box size but a particle mass that is eight times larger (80-512) or smaller (80-2048).  The $\delta_{\rm m}$ distribution from C16 is displayed as the dashed black histogram in the upper panel.}
    \label{fig:CoSLA-ResComp}
\end{figure}

\begin{figure}
    \centering
    \includegraphics{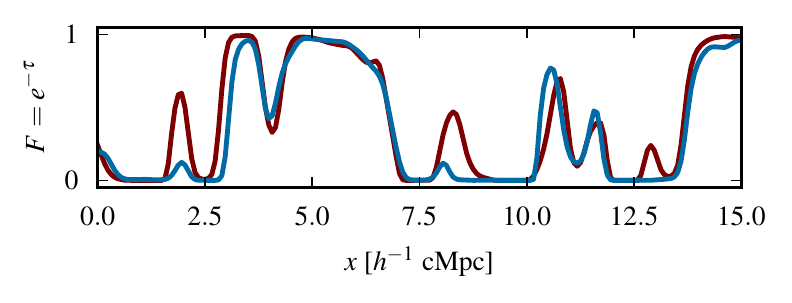}
    \vspace{-4mm}
    \caption{Simulated CoSLAs from the same segment drawn from the 80-512 (blue) and 80-2048 (dark red) models. This segment corresponds to $\delta_{\tau_{\rm eff}}=3.90$ in 80-512 but only $\delta_{\tau_{\rm eff}}=3.67$ in 80-2048, due to the increased transmission from the better resolved underdense gas in the higher resolution model.} 
    \label{fig:SpecResComp}
\end{figure}

The effect of an increase in box size on the $\delta_m$ and \deltaTeff\ distributions for CoSLAs is displayed in Figure~\ref{fig:CoSLA-BoxComp}.  The $\delta_m$ distribution in the largest volume simulation (160-2048) extends toward higher values as a consequence of the rarer, more massive systems present in the larger volume.  The number of CoSLAs per unit volume increases by $77$ per cent in the 160-2048 simulation (see Table~\ref{tab:ConvSims}) relative to the fiducial 80-1024 (Q\lya) model, although both the 80-1024 and 160-2048 models share similar median values of $\delta_{\rm m}=0.40\pm0.02$
and $\delta_{\rm m}=0.48\pm0.05$, respectively.  The corresponding \deltaTeff\ distribution displays similar behaviour, with a tail that extends to larger values of \deltaTeff\ in the 160-2048 model.  The effect of the $\sim 1000$ times larger volume used by C16 therefore likely explains some of the differences relative to this work.  The CoSLAs selected in C16 (shown by the dashed black histogram in the upper panel of Figure~\ref{fig:CoSLA-BoxComp}) probe rarer, more massive systems that are not present in our small, higher resolution models.  We are unable to reliably assess how effectively CoSLAs probe overdense $15^{3}h^{-3}\rm\,cMpc^{3}$ volumes with $\delta_{\rm m} > 1$ using our fiducial $80^{3}h^{-3}\rm\, cMpc^{-3}$ simulation box.  However, this still does not fully explain the greater incidence of CoSLAs at overdensites of $\delta_{\rm m}<0.5$ within the hydrodynamical simulations compared to C16, even when using Q\lya\ models that underpredict the number of CoSLAs at $\delta_{\rm m}\leq 0.4$ due to missing LLSs.  A possible explanation is that the lack of high $\tau_{\rm eff}$ systems at $\delta_{\rm m}$ is even more pronounced in much larger collisionless simulations, due to unresolved high column density absorption from small scale structure, but the exact reason for this difference remains unclear.

Mass resolution also has an important -- but more subtle -- effect on the typical \deltaTeff\ and $\delta_m$ associated with CoSLAs. The top panel of Figure~\ref{fig:CoSLA-ResComp} highlights how the $\delta_{\rm m}$ distribution of CoSLAs is shifted toward lower values for the 80-512 simulation when compared to the two higher resolution models. The median $\delta_{\rm m}=0.38\pm0.01$ for the 80-512 model, compared to median values of $\delta_{\rm m}=0.40\pm0.02$ and $\delta_{\rm m}=0.42\pm0.04$ for 80-1024 (Q\lya) and 80-2048, respectively.  However, the average number of CoSLAs per unit volume also decreases by a factor $\sim 2$ when increasing the mass resolution by a factor of 8 (see Table~\ref{tab:ConvSims}), suggesting this quantity is not yet converged with mass resolution.  

The lower panel of Figure~\ref{fig:CoSLA-ResComp} elucidates the origin of this behaviour: at higher mass resolution the high \deltaTeff\ tail of the CoSLA distribution is truncated in comparision with the lower resolution runs.  This is a consequence of the failure of the lower resolution runs to correctly resolve the structure of the \lya\ absorption, particularly in underdense regions, which are too opaque in low mass resolution models \citep{Bolton2009ResolvingSimulations}.  The effect of this poorly resolved underdense gas is to increase the number of volumes on $15h^{-1}\rm\,cMpc$ scales that are associated with segments which exceed the CoSLA identification threshold of $\delta_{\tau_{\rm eff}}=3.5$.  An illustration of this effect is shown in Figure~\ref{fig:SpecResComp}, where \lya\ absorption from a randomly selected segment in the 80-2048 simulation (dark red) is compared directly to the 80-512 simulation (blue).  This implies that -- in addition to not capturing absorption from high column density, self-shielded absorbers -- low mass resolution models that do not adequately resolve the structure of the \lya\ forest on small scales will overpredict the incidence of CoSLAs above a fixed \deltaTeff\ threshold.  

\section{Characterising protoclusters in \lya\ absorption}\label{ssec:PCFrac}

We now address the question of how protoclusters, as opposed to overdensities, are characterised by \lya\ absorption.  Whilst -- observationally -- protoclusters are identified as overdense regions at $z\sim2$, when using simulations we have access to \emph{a priori} knowledge of which structures will collapse to form a cluster along with the associated cluster mass. We now use this simulation based \lq\lq true" definition of a protocluster to examine the opacity of protoclusters at $z=2.4$ on $15\mpc$ scales.  We do not expect variations in the numerical methodology used by Sherwood, EAGLE and Illustris to significantly change these results, so from this point onward we focus on analysing only the Sherwood simulation.

\begin{figure}
	\centering
	\includegraphics[width=\columnwidth]{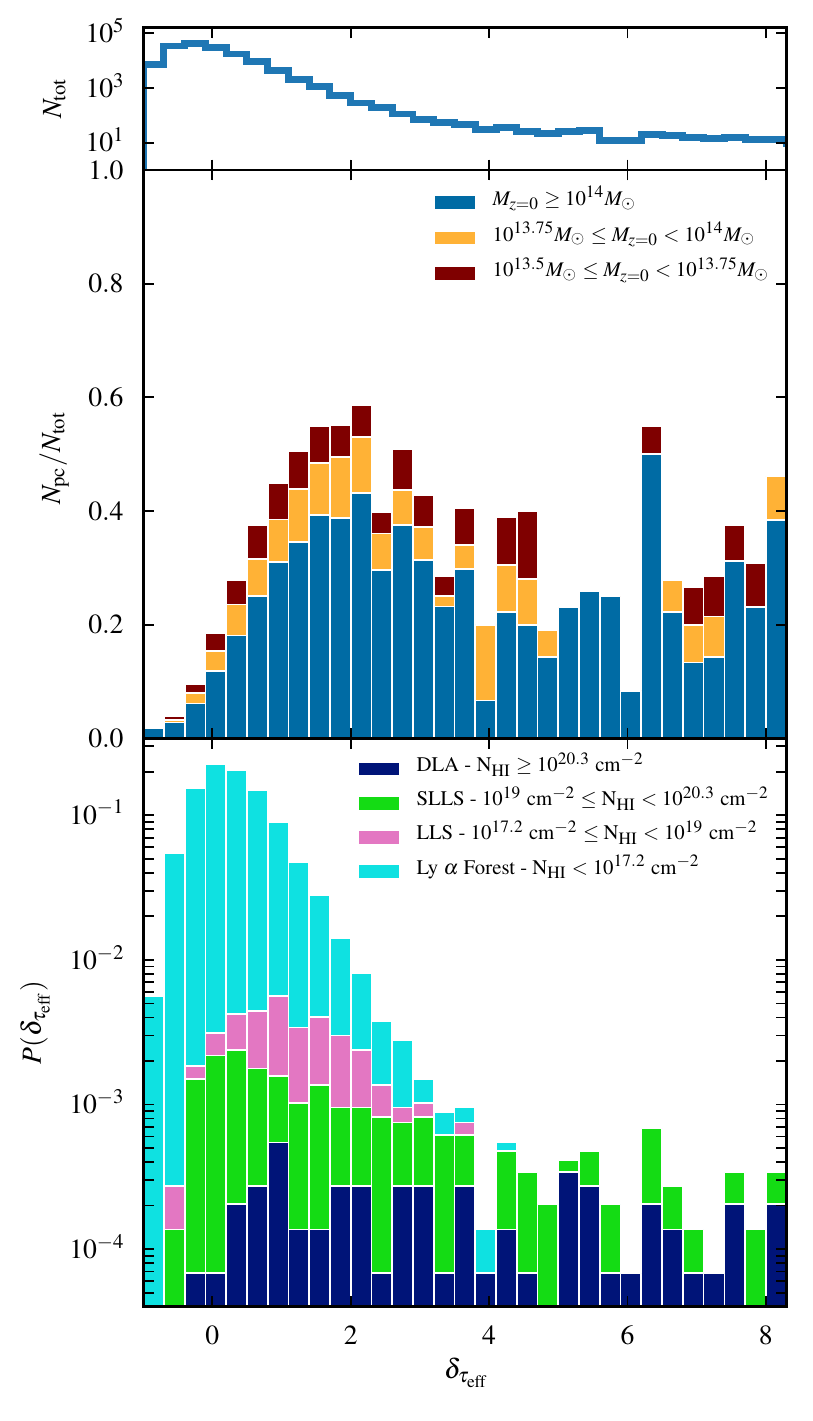}
	\vspace{-0.8cm}
    \caption{Top: Total number of $15h^{-1}\rm\,cMpc$ segments in bins of $\Delta \delta_{\tau_{\rm eff}}=0.3$. 
    Middle: The fraction of $15h^{-1}\rm\,cMpc$ segments in bins of $\Delta \delta_{\tau_{\rm eff}}=0.3$ associated with protoclusters or protogroups.  The colour of each stacked bar indicates the mass of the protocluster or protogroup associated with the absorption. 
    Bottom: probability distribution for all $15h^{-1}\rm\,cMpc$ segments that are associated with protoclusters. The colour of each stacked bar indicates the maximum column density within each segment.}
    \label{fig:PCFrac}
\end{figure}

\begin{figure*}
    \includegraphics[width=\textwidth]{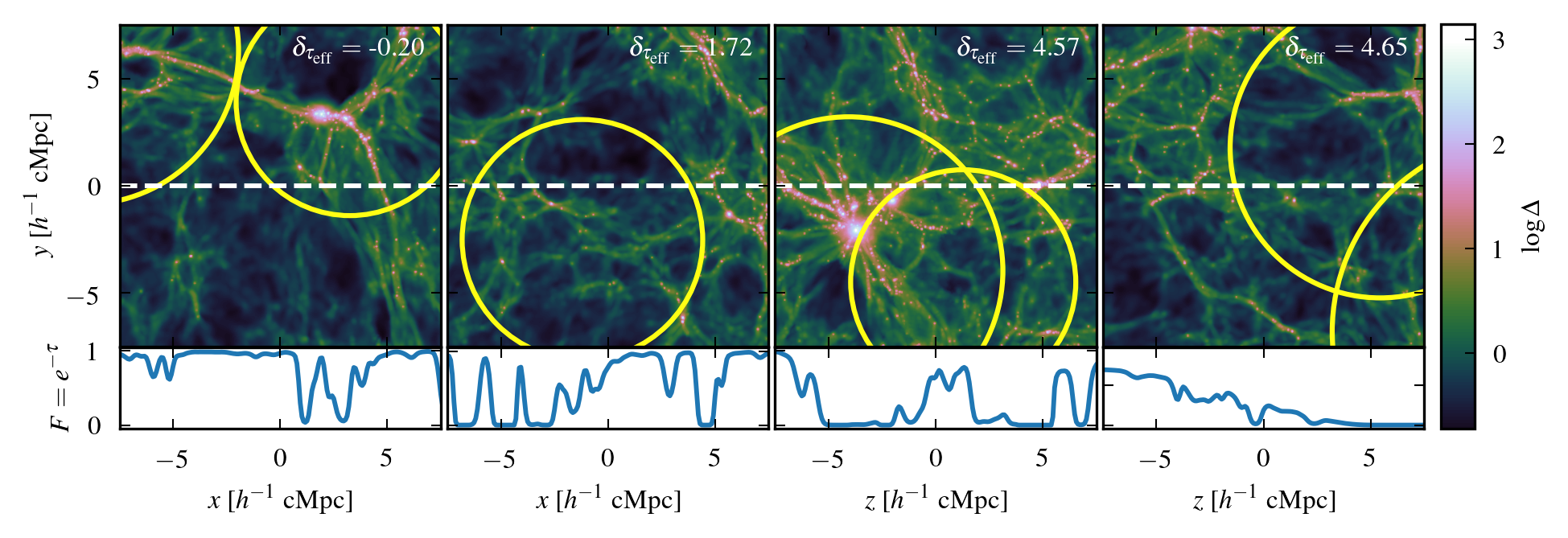}
    \vspace{-0.8cm}
    \caption{Example of four different $15h^{-1}\rm\,cMpc$ segments that pass through protoclusters in Sherwood. The top panels show projected maps of the normalised gas density in a $15h^{-2}\rm\,cMpc^{2}$ slice with a projection depth of $1\mpc$. The white dashed line shows the direction in which the \lya\ absorption spectrum was extracted,  while the yellow circles indicate the cross section of all protoclusters that intersect with the slice. The bottom panels show the corresponding mock \lya\ absorption spectra. From left to right, each segment corresponds to: a low opacity segment with $\delta_{\tau_{\rm eff}}=-0.20$ containing a maximum \HI\ column density $N_{\rm HI}=10^{13.8}~\textrm{cm}^{-2}$, a higher opacity segment with $\delta_{\tau_{\rm eff}}=1.72$ and maximum $N_{\rm HI}=10^{14.8}~\textrm{cm}^{-2}$, a CoSLA with $\delta_{\tau_{\rm eff}}=4.57$ and maximum  $N_{\rm HI}=10^{16.9}~\textrm{cm}^{-2}$, and a DLA with $\delta_{\tau_{\rm eff}}=4.65$ and maximum $N_{\rm HI}=10^{20.5}~\textrm{cm}^{-2}$.}
    \label{fig:PCTeffEx}
\end{figure*}

In Figure~\ref{fig:PCFrac}, the fraction of all $15h^{-1}\rm\,cMpc$ segments in bins of $\Delta$\deltaTeff~$=0.3$ that are associated\footnote{A $15\mpc$ segment is associated with a protocluster or protogroup if at least one third of the segment passes within \rnf\ of any protocluster/group.} with protoclusters (blue shading) or protogroups (orange and brown shading) are displayed in the central panel.  In the event that a system is associated with more than one protocluster, it is associated with the most massive one only.  The corresponding number of $15\mpc$ segments in each bin is shown in the upper panel, with most segments close to the mean, $\delta_{\tau_{\rm eff}}=0$.

Note that in almost all bins, the majority of the ``associated" segments are in the protoclusters with $M_{\rm z=0}\geq 10^{14}\,M_{\odot}$.  This in part reflects the fact that the protoclusters occupy a larger fraction of the simulation volume relative to the protogroups\footnote{Protoclusters occupy 7.7 per cent of the Sherwood simulation volume at $z=2.4$, whereas the large and small protogroups together only occupy 5.0 per cent.)}.   Typically fewer than 10 per cent of segments with $\delta_{\tau_{\rm eff}} \lesssim 0$ are associated with protoclusters, increasing to $\sim 40$ per cent at $\delta_{\tau_{\rm eff}}=2$. At $\delta_{\tau_{\rm eff}}\gtrsim2$, however, this fraction declines, where on average only $\sim25$ per cent of segments are associated with protoclusters. Figure~\ref{fig:PCFrac} therefore demonstrates that for $\delta_{\tau_{\rm eff}}<2$ there is a positive, albeit weak, correlation between \deltaTeff\ and the likelihood of the associated volume collapsing to form a cluster by $z=0$.

In the lower panel of Fig.~\ref{fig:PCFrac}, the probability distribution for $15h^{-1}\rm\,cMpc$ segments residing within the protoclusters is displayed as a function of \deltaTeff, with the shading indicating the fraction of segments in each bin that contains a given maximum \HI\ column density.  The majority of segments passing through protoclusters only contain \lya\ forest absorption systems and have $\delta_{\tau_\textrm{eff}}\sim0$: 28 per cent of the segments which are associated with protoclusters are \lya\ forest with $\delta_{\tau_{\rm eff}}<0$, and 84 per cent are entirely \lya\ forest with $\delta_{\tau_{\rm eff}}<1$. Thus, while the majority of segments with \deltaTeff$<1$ do not correspond to protoclusters, but rather to ``field'' regions, the majority of 
segments associated \emph{with} protoclusters lie in the same \deltaTeff\ range and share identical characteristics.   This reflects the fact that protoclusters at $z\simeq 2.4$ are extended over large scales \citep{Muldrew2015WhatProtoclusters};
this means that it is very difficult -- or impossible -- to disentangle most protocluster sight lines from the field on $15\mpc$ scales.

At larger values of \deltaTeff\ the segments associated with protoclusters become dominated by SLLSs and DLAs. At $\delta_{\tau_{\rm eff}}>2.9$, the majority of segments associated with protocluster contain damped systems, and 96 per cent of the segments with $\delta_{\tau_{\rm eff}} \gtrsim 3.5$ (i.e. the threshold for CoSLAs) that are associated with protoclusters arise from high density gas producing damped \lya\ absorption. At $\delta_{\tau_{\rm eff}}>3.5$ SLLSs and DLAs are the dominant probe of gas that forms clusters by $z=0$, but  -- critically for protocluster selection by \lya\ absorption -- the middle panel of Figure.~\ref{fig:PCFrac} shows that these are not uniquely associated with protoclusters at $z\simeq 2.4$, but are more likely to be in the field.

Figure~\ref{fig:PCTeffEx} displays selected examples of segments that pass through protoclusters that cover the full range of \deltaTeff\ observed in the simulation. Most sight lines passing through protoclusters with $\delta_{\tau_{\rm eff}}<2$ are exemplified by the two left-hand panels in Figure~\ref{fig:PCTeffEx}.  In these cases the sight line passes through two filamentary structures perpendicular to the sight line (left panel, $\delta_{\tau_{\rm eff}}=-0.20$) or through a small region of overdense gas within a protocluster (second left panel, $\delta_{\tau_{\rm eff}}=1.72$), but in both cases do not exhibit extended regions of saturated \lya\ absorption.   

By contrast, in the two right-hand panels of Figure~\ref{fig:PCTeffEx}, two examples of segments that pass through protoclusters and have $\delta_{\tau_{\rm eff}}>3.5$ (i.e. the CoSLA threshold) are displayed.  The right-most panel ($\delta_{\tau_{\rm eff}}=4.65$) displays strong absorption due to the presence of an extended damping wing arising from a DLA; there is no large scale gas overdensity present along the sight line.  The segment displayed on the second right ($\delta_{\tau_{\rm eff}}=4.57$) has no damped absorbers, but instead the absorption arises from haloes and a filament that are aligned with the sight line and are associated with two intersecting protoclusters.  The high value of \deltaTeff\ in this case results from extended \lya\ forest absorption.  This implies that it is the orientation of dense gas with respect to the sight line that has the greatest impact on the value of \deltaTeff, rather than either the overdensity $\delta_{\rm m}$ (see also Figure~\ref{fig:LowMassCont}) or the presence of a protocluster.

\section{Selecting protoclusters with line of sight \lya\ absorption}\label{sec:CompCont}

\begin{figure}
    \includegraphics[width=\columnwidth]{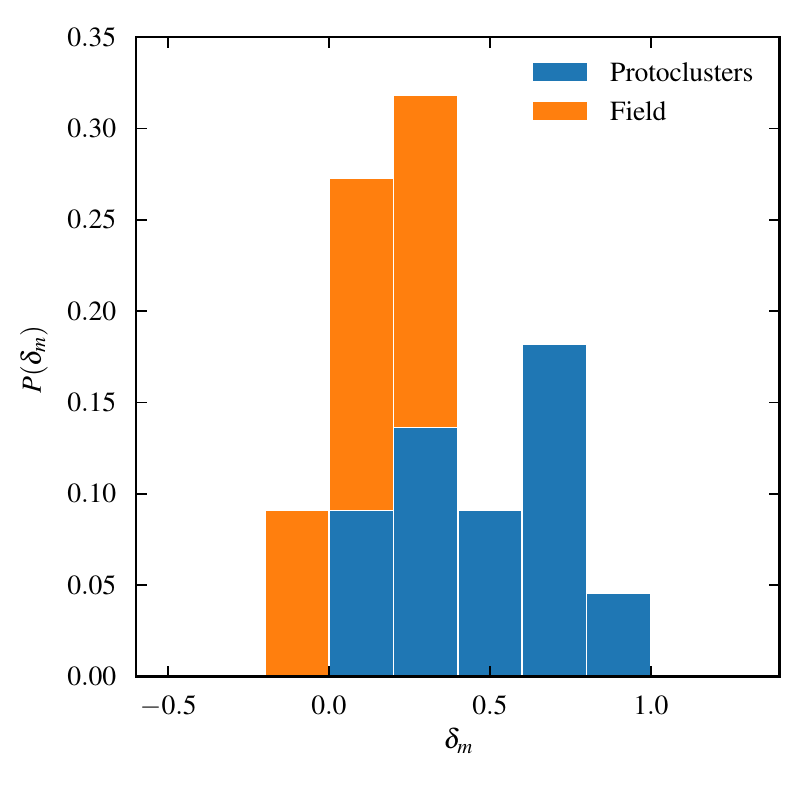}
    \vspace{-0.8cm}
    \caption{The probability distribution of CoSLAS identified in Sherwood as a function of $\delta_\textrm{m}$, coloured by whether the CoSLA is associated with a protocluster (12; blue) or associated with the field (10; orange).}
    \label{fig:PCVsFldPDF}
\end{figure}

Now that we have characterised the \lya\ absorption associated with protoclusters, we finally turn to examine the effectiveness with which one can select protocluster regions using line of sight \lya\ absorption.  We have established that most of the sight lines that pass through protoclusters with $M_{\rm z=0}\sim 10^{14}\,M_{\odot}$ exhibit low values of \deltaTeff, and it is impossible to distinguish these from the field.  On the other hand, for large values of \deltaTeff\ most of the segments associated with protoclusters are the result of damped absorption systems with $N_{\rm HI}>10^{19}~\textrm{cm}^{-2}$, but these systems do not uniquely trace protoclusters. 

In addition, in the simulations there are a small subset of high opacity segments within protoclusters that do not exhibit damped absorption (e.g. the example CoSLA shown in the second from right panel in Figure~\ref{fig:PCTeffEx}).   This is demonstrated further in Figure~\ref{fig:PCVsFldPDF}, which shows the distribution of CoSLAs in Sherwood split into two groups as a function of $\delta_{\rm m}$:  those that are associated with a protocluster and those that are not.  We find 55 per cent of CoSLAs are associated with protoclusters with a median $\delta_m = 0.4\pm0.1$, in comparison to CoSLAs that probe the field, with median $\delta_m = 0.06\pm0.05$.  This indicates that, while the two populations cannot be separated by their \lya\ absorption spectra alone, approximately half of CoSLAs -- which in general arise from the alignment of overdense structure along the line of sight -- are indeed associated with protoclusters.  Thus we find that CoSLAs are not a unique tracer of protoclusters, even when assuming perfect removal of damped systems from the mock data.

\begin{figure}
	\includegraphics[width=\columnwidth]{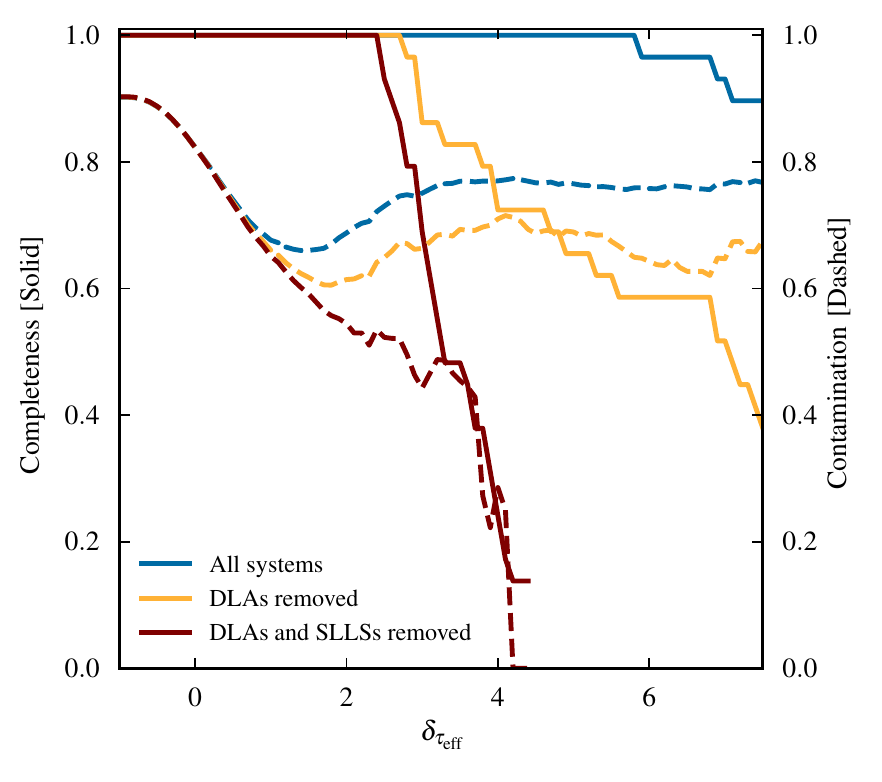}
		\vspace{-0.8cm}
    \caption{Contamination (dashed) and \lq\lq completeness" (solid) for protoclusters (with $M_{\rm z=0}\geq 10^{14}M_{\odot}$) when selecting all segments above a fixed threshold of  \deltaTeff\ on $15h^{-1}\rm\,cMpc$ scales (see text for details). The blue curves correspond to the case where all absorption is considered, orange curves to the case where sight lines containing DLAs are removed, and brown curves to the case where SLLSs and DLAs are removed. Results are displayed only where there is more than one $15h^{-1}\rm\,cMpc$ segment in each bin.} 
    \label{fig:ComContPs13+SS}
\end{figure}

\begin{figure}
    \includegraphics[width=\columnwidth]{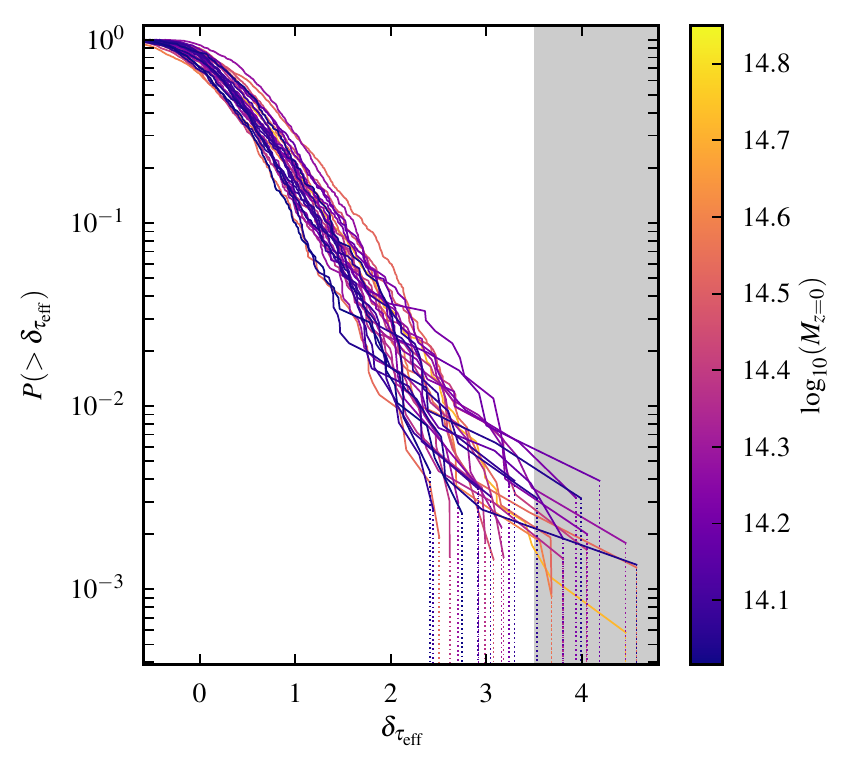}
    \vspace{-0.8cm}
    \caption{Reverse cumulative distribution functions, $P(>\delta_{\tau_\textrm{eff}})$, for all $15h^{-1}\rm\,cMpc$ segments that are associated with each of the 29 protoclusters in Sherwood.  Each line is coloured according to the mass of the cluster at $z=0$. The grey shaded region indicates the area above the CoSLA selection threshold of $\delta_{\tau_\textrm{eff}} > 3.5$.}
    \label{fig:SurFunc}
\end{figure}

\begin{figure*}
    \centering
    \includegraphics{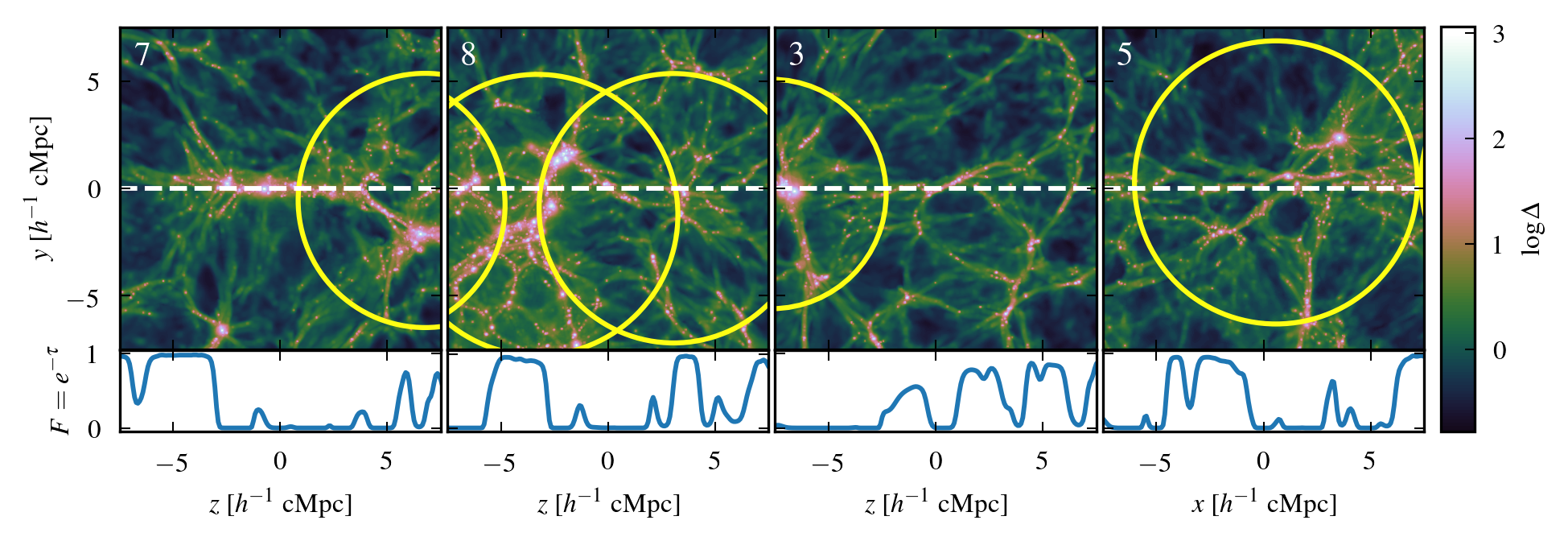}
    \vspace{-5mm}
    \caption{Example of four different CoSLAs that pass through protoclusters in Sherwood. The top panels show projected maps of the normalised gas density in a $15h^{-2}\rm\,cMpc^{2}$ slice with a projection depth of $1\mpc$. The white dashed line shows the direction in which the \lya\ absorption spectrum was extracted. The bottom panels show the corresponding mock \lya\ absorption spectra. The numbers in the upper left of each panel correspond to the CoSLAs listed in Table~\ref{tab:PC_CoSLAs}. From left to right, the maximum \HI\ column density associated with each CoSLA is $N_{\rm HI}=10^{16.5}~\textrm{cm}^{-2}$, $10^{14.9}~\textrm{cm}^{-2}$, $10^{19.0}~\textrm{cm}^{-2}$ and $10^{14.7}~\textrm{cm}^{-2}$.}
    \label{fig:PCCosLARenders}
\end{figure*}

The effective optical depth threshold that defines a CoSLA as a segment exhibiting $\delta_{\tau_{\rm eff}}>3.5$ on $15h^{-1}\rm\,cMpc$ scales is to some extent arbitrary.  We may therefore consider whether a fixed threshold in $\delta_{\tau_{\rm eff}}$ can be chosen that minimises the contamination within a sample of protoclusters selected from the mock spectra.  We define the contamination as  $N(\geq \delta_{\tau_{\rm eff}})/N_{\rm tot}(\geq \delta_{\tau_{\rm eff}})$, where $N(\geq \delta_{\tau_{\rm eff}})$ is the number of $15h^{-1}\rm\,cMpc$ segments above the \deltaTeff\ threshold that do not pass through at least 5$h^{-1}\rm\,cMpc$ of a protocluster volume, and $N_{\rm tot}(\geq \delta_{\tau_{\rm eff}})$ is the total number of $15h^{-1}\rm\,cMpc$ segments above the \deltaTeff\ threshold in the entire simulation volume. We also define the sample \lq\lq completeness" for a given \deltaTeff\ threshold as the fraction of protoclusters with at least one sight line that exhibits absorption on $15h^{-1}\rm\,cMpc$ scales above the  $\delta_{\tau_{\rm eff}}$ threshold. This is defined as $N_{\rm pc}(\geq \delta_{\tau_{\rm eff}})/N_{\rm pc, tot}$, where $N_{\rm pc}(\geq \delta_{\tau_{\rm eff}})$ is the number of unique protoclusters probed by $15h^{-1}\rm\,cMpc$ segments above the \deltaTeff\ threshold, and $N_{\rm pc,tot}$ is the total number of protoclusters in the simulation (i.e. 29 for Sherwood, see Table~\ref{tab:Sims}).
 
These values are plotted as a function of the \deltaTeff\ threshold in Figure~\ref{fig:ComContPs13+SS} for three different cases: where absorption in all the sight lines is considered (blue curves),  where sight lines containing DLAs are removed from the sample (orange curves), and the case where both SLLSs and DLAs are removed (brown curves).  When considering all sight lines, the protocluster sample remains 100 per cent complete using a threshold up to $\delta_{\tau_\textrm{eff}} \sim 6$ because all protoclusters contain DLA systems that produce such high $\delta_{\tau_\textrm{eff}}$. The contamination also remains very high at $\sim 75$ per cent, which reflects the fact that the damped \lya\ absorbers do not uniquely probe protocluster environments.  If we instead assume perfect removal of all sight lines containing DLAs, the contamination rate falls  slightly to around 65 per cent for the segments with the strongest absorption.  Finally, removing all the $15h^{-1}\rm\,cMpc$ segments containing SLLSs as well as DLAs causes the contamination to drop to zero at $\delta_{\tau_\textrm{eff}}=4.1$, i.e. all the remaining segments with this \deltaTeff\ uniquely trace protocluster gas.  Consequently, if we assume perfect removal of damped absorption systems, it is possible to obtain a perfectly clean sample of protoclusters by applying a sufficiently high threshold in $\delta_{\tau_\textrm{eff}}$.  

This clean sample of protoclusters is, however, highly incomplete as not all protoclusters will exhibit strong coherent \lya\ absorption on $15h^{-1}\rm\,cMpc$ scales.  This can be seen in Figure~\ref{fig:ComContPs13+SS}, where the drop in contamination is accompanied by the completeness falling to only $17$ per cent. This means that less than a fifth\footnote{The completeness is likely to be strongly dependent on the protocluster mass. Since we do not probe massive protoclusters -- due to our small box size -- this completeness is likely to be a lower limit.} of protoclusters have any sight lines with $\delta_{\tau_{\rm eff}}>4.1$. This is further exemplified in Figure~\ref{fig:SurFunc}, which shows the reverse cumulative distribution function of \deltaTeff\ for each of the 29 protoclusters in Sherwood. The colour of each line corresponds to the mass of the resulting cluster at $z=0$. This demonstrates that high $\delta_{\tau_{\rm eff}}$ segments that pass through protoclusters are rare: less than 0.1 per cent of segments that pass through $M_{\rm z=0}\sim 10^{14}\rm\,M_{\odot}$ protoclusters have $\delta_{\tau_{\rm eff}}>3.5$ -- corresponding to the CoSLA theshold defined by C16, shown by the grey shading. 

All the CoSLAs associated with protoclusters in Sherwood are listed in detail in Table~\ref{tab:PC_CoSLAs}: a total of 14 unique protoclusters over a mass range of $10^{14.0}$--$10^{14.7}~\msol$ are probed by 12 unique CoSLAs, and 42 per cent of the CoSLAs associated with protoclusters pass through more than one protocluster. A sample of four of the CoSLAS associated with protoclusters in Sherwood are displayed in Figure~\ref{fig:PCCosLARenders}.  We observe again that typically there is an alignment of structure along the sight line that causes the coherent \lya\ absorption.  Orientation of structure to the line of sight, rather than association with a protocluster of a given mass or an overdensity, appears to be a critical factor that determines the extended nature of the \lya\ absorption.   In this context, we briefly note that \citet{Finley2014AstrophysicsProtocluster} argued for the detection of an intergalactic filament  based on observations of multiple LLSs and SLLSs with $N_{\rm HI}>10^{18}\rm\,cm^{-2}$ along two closely separated quasar sight lines at $z=2.69$.  The seven strong \HI\ absorption systems observed by \citet{Finley2014AstrophysicsProtocluster} span $\sim 1700\rm\,km\,s^{-1}$, corresponding to $16.6h^{-1}\rm\,cMpc$ at $z=2.69$. While  \citet{Finley2014AstrophysicsProtocluster} could not definitively rule out association of the \HI\ absorbers with a protocluster, their favoured interpretation is broadly consistent with our analysis.   Taken a step further, this suggests that CoSLAs may in fact be a tracer of extended filamentary structure in the early Universe. An intriguing possibility is that CoSLAs, and their association (if any) with galaxies and/or metal absorption lines, may therefore provide a route to identifying and studying filamentary environments at $z>2$.

\begin{table}
    \centering
        \caption{All CoSLAs (defined as $15h^{-1}\rm\,cMpc$ segments with $\delta_{\tau_{\rm eff}}>3.5$ that do not contain  a damped \lya\ absorber) in Sherwood that are associated with protoclusters.  A total of 14 protoclusters are probed by 12 unique CoSLAs. Note that each CoSLA may probe multiple protoclusters.}
    \begin{tabular}{c|c|c|c}
        \hline
        CoSLA & \deltaTeff & $\delta_m$  & Protocluster $M_{z=0}$ \\
        & & & $\left[\log_{10}(\msol)\right]$\\
        \hline
        1 & 4.57 & 0.56 & 14.54, 14.04\\
        2 & 4.46 & 0.80 & 14.72, 14.28\\
        3 & 4.18 & 0.38 & 14.19\\
        4 & 4.05 & 0.37 & 14.25\\
        5 & 4.05 & 0.42 & 14.41\\
        6 & 3.99 & 0.12 & 14.05 \\
        7 & 3.93 & 0.35 & 14.22\\
        8 & 3.80 & 0.72 & 14.35, 14.18\\
        9 & 3.68 & 0.98 & 14.59, 14.53\\
        10 & 3.67 & 0.70 & 14.72, 14.53\\
        11 & 3.56 & 0.62 & 14.59\\
        12 & 3.53 & 0.12 & 14.06\\
        \hline
    \end{tabular}

    \label{tab:PC_CoSLAs}
\end{table}

\section{Conclusions}\label{sec:Conclusion}

In this work we have used state of the art hydrodyamical simulations from the Sherwood \citep{Bolton2017The5}, EAGLE \citep{Schaye2015TheEnvironments, Crain2015TheVariations, McAlpine2016TheCatalogues} and Illustris \citep{Vogelsberger2014IntroducingUniverse, Nelson2015TheRelease} projects to examine the signature of protoclusters observed in \lya\ absorption at $z\simeq 2.4$.  Building upon earlier work using low resolution collisionless dark matter simulations \citep[e.g.][]{Stark2015ProtoclusterMaps,Cai2016MAppingMethodology}, here we use models that resolve small scale structure in the IGM, correctly reproduce the incidence of \HI\ absorption systems (including high column densities that are self-shielded to Lyman continuum photons), and track the formation of structure to $z=0$.  

We examine the impact of small scale gas structure on the signature of large scale overdensities on $15h^{-1}\rm\,cMpc$ scales, finding that the simulation mass resolution required for resolving \lya\ absorption on small scales is also a requirement for correctly modelling the average $\tau_{\rm eff}$ on $15\mpc$ scales. This is necessary for capturing the incidence of \HI\ absorption systems over a wide range of column densities, including damped systems and LLSs.  At the same time, however, adequate mass resolution is necessary for correctly capturing the opacity of underdense regions in the IGM.  A mass resolution that is too low will overpredict the typical \lya\ effective optical depth on $15\mpc$ scales.

We furthermore assess the prevalance of coherent \lya\ absorption within protoclusters at high redshift.  Our main conclusions may be summarised as follows:

\begin{itemize}

    \item We confirm there is a weak correlation between the mass overdensity, $\delta_m$, and the effective optical depth relative to the mean, \deltaTeff, on a $15~\mpc$ scales, in the simulations, although there is a large amount of scatter that, particularly at large values of \deltaTeff, means it is not possible to uniquely identify large scale overdensities with strong \lya\ absorption.  This remains true even if first removing all damped \lya\ absorption systems that arise from dense, neutral gas on small scales. 
    
    \item We examine the properties of coherently strong intergalactic \lya\ absorption systems (CoSLAs) in the simulations.  CoSLAs -- defined by C16 as regions on $15h^{-1}\rm\,cMpc$ scales with $4.5$ times the average \lya\ effective optical depth after excluding damped absorption systems --  are rare objects, only accounting for 0.1 per cent of all $15\mpc$ spectral segments drawn from the models.  They probe a wide range in mass overdensity, $\delta_{\rm m}$, including underdense regions on $15h^{-1}\rm\,cMpc$ scales, and so do not uniquely trace significant mass overdensities.
   
    \item Protoclusters with $M_{\rm z=0}\simeq 10^{14}\,M_{\odot}$ exhibit a broad range of signatures in \lya\ absorption, with  $\delta_{\tau_{\rm eff}}$ ranging from $-0.5$ to $>8$. However, the vast majority (84 per cent) of sight lines passing through what we define as protoclusters in the simulations contain only low column density \lya\ forest absorption and have $\delta_{\tau_{\rm eff}}<1$. This signature is identical to the field, so the majority of sight lines through protoclusters do not bear a telltale signature in \lya\ absorption. Most sight lines with high $\delta_{\tau_{\rm eff}}$ are a result of passing through a damped absorption system.  A small subset of sight lines through protoclusters do, however, exhibit high $\delta_{\tau_{\rm eff}}$ due to coherently strong intergalactic \lya\ absorption systems i.e. CoSLAs.  LLSs and high column density \lya\ forest absorbers are typically responsible for this absorption.
    
    \item Assuming the perfect removal of damped \lya\ absorption systems, CoSLAs are a good but non-unique probe of protoclusters at $z\simeq 2.4$.  In the Sherwood simulation, approximately half of CoSLAs with $\delta_{\tau_{\rm eff}}>3.5$ trace protoclusters with $10^{14} \leq M_{\rm z=0}/M_{\odot} \leq 10^{14.7}$.  The other 46 per cent of CoSLAs arise from LLSs that are aligned along the line of sight.
    
    \item We find that threshold of $\delta_{\tau_{\rm eff}}>4.1$ -- corresponding to regions on $15h^{-1}\rm\,cMpc$ scales with $5.1$ times the average \lya\ effective optical depth after excluding damped absorption systems -- enables us to select a completely pure sample of protoclusters from simulated spectra.  However, CoSLAs are rare within the volumes that protoclusters occupy: less than $0.1$ per cent per cent of sight lines that pass through at least $5h^{-1}\rm\,cMpc$ of a protocluster volume exhibit $\delta_{\tau_{\rm eff}}>4.1$, excluding absorption caused by damped systems.  This means that any sample of protoclusters selected with the CoSLA technique will be incomplete.  We stress, however, that throughout this work we are limited by the box size of the hydrodynamical simulations. In particular, there are no $10^{15}~\msol$ cluster progenitors in any of the models considered here; it is possible that more massive protoclusters have a stronger association with CoSLAs.
\end{itemize}

\noindent
Finally, we note that visual inspection of CoSLAs suggests that coherent \lya\ absorption typically selects structure orientated along the line of sight to the observer, regardless of whether or not this is associated with a protocluster.   With the advent of large spectroscopic QSO surveys such DESI \citep{Vargas-Magana2019UnravelingDESI} and WEAVE-QSO \citep{Pieri2016WEAVE-QSO:Telescope} in the next few years, further investigation of the potential of CoSLAs for identifying intergalactic filaments in the high redshift Universe may be a worthwhile endeavour.   

\section*{Acknowledgements}

We thank Zheng Cai for comments on a draft version of this paper.  The Sherwood simulations were performed with supercomputer time awarded by the Partnership for Advanced Computing in Europe (PRACE) 8th Call. We acknowledge PRACE for awarding us access to the Curie supercomputer, based in France at the Tre Grand Centre de Calcul (TGCC). This work also made use of the DiRAC Data Analytic system at the University of Cambridge, operated by the University of Cambridge High Performance Computing Service on behalf of the STFC DiRAC HPC Facility (www.dirac.ac.uk). This equipment was funded by BIS National E-infrastructure capital grant (ST/K001590/1), STFC capital grants ST/H008861/1 and ST/H00887X/1, and STFC DiRAC Operations grant ST/K00333X/1. DiRAC is part of the National E-Infrastructure.  We thank Volker Springel for making \textsc{P-GADGET-3} available, and Simeon Bird for helpful advice regarding the incidence of DLAs in Illustris.  We acknowledge the Virgo Consortium for making their simulation data available.  The \textsc{EAGLE} simulations were performed using the DiRAC-2 facility at Durham, managed by the ICC, and the PRACE facility Curie in France at TGCC.  We also thank the Illustris collaboration for making their data publicly available.  JSAM is supported by an STFC postgraduate studentship.  JSB acknowledges the support of a Royal Society University Research Fellowship.




\bibliographystyle{mnras}
\bibliography{Mendeley}




\appendix


\bsp	
\label{lastpage}
\end{document}